\documentclass[usenatbib]{mnras}

\usepackage{graphicx,rotating}
\usepackage{amssymb,amsmath,pdflscape}
\usepackage{natbib}
\usepackage{subeqnarray}
\usepackage{mathtools}
\usepackage{cases}
\usepackage[utf8x]{inputenc}
\usepackage{auto-pst-pdf}
\usepackage{hyperref}
\usepackage{cleveref}

\newcommand{\pc}{p_{\rm c}}
\newcommand{\rhoc}{\rho_{\rm c}}
\newcommand{\psic}{\Psi_{\rm c}}

\newcommand{\const}{{\rm const.}}

\newcommand{\xij}{{x_{i,j}}}
\newcommand{\cij}{c_{i,j}}
\newcommand{\qij}{q_{i,j}}
\newcommand{\fij}{f_{i,j}}
\newcommand{\kji}{k_{j,i}}
\newcommand{\kij}{k_{i,j}}
\newcommand{\lambdaij}{\lambda_{i,j}}

\newcommand{\epsijprim}{{\epsilon_{i,j}'}}

\newcommand{\rhol}{\rho_{\cal L}}
\newcommand{\epsl}{\epsilon_{\cal L}}
\newcommand{\epsj}{\epsilon_j}
\newcommand{\epsi}{\epsilon_i}
\newcommand{\barei}{\bar{\epsilon}_i}
\newcommand{\barej}{\bar{\epsilon}_j}
\newcommand{\bareone}{\bar{\epsilon}_1}

\newcommand{\barell}{\bar{\epsilon}}

\newcommand{\trho}{\tilde{\rho}}
\newcommand{\tOmega}{\tilde{\Omega}}

\begin{document}

\title[]{Nested spheroidal figures of equilibrium\\ II. Generalization to ${\cal L}$ layers}

\author[J.-M. Hur\'e]
       {J.-M. Hur\'e$^{1,2}$\thanks{E-mail:jean-marc.hure@u-bordeaux.fr}\\
$^{1}$Univ. Bordeaux, LAB, UMR 5804, F-33615, Pessac, France\\
$^{2}$CNRS, LAB, UMR 5804, F-33615, Pessac, France}

\date{Received ??? / Accepted ???}
 
\pagerange{\pageref{firstpage}--\pageref{lastpage}} \pubyear{???}

\maketitle

\label{firstpage}

\begin{abstract}
  We present a vectorial formalism to determine the approximate solutions to the problem of a composite body made of ${\cal L}$ homogeneous, rigidly rotating layers bounded by spheroidal surfaces. The method is based on the $1$st-order expansion of the gravitational potential over confocal parameters, thereby generalizing the method described in Paper I for ${\cal L}=2$. For a given relative geometry of the ellipses and a given set of mass-density jumps at the interfaces, the sequence of rotation rates and interface pressures is obtained analytically by recursion. A wide range of equilibria result when layers rotate in an asynchronous manner, although configurations with a negative oblateness gradient are more favorable. In contrast, states of global rotation (all layers move at the same rate), found by solving a linear system of ${\cal L}-1$ equations, are much more constrained. In this case, we mathematically demonstrate that confocal and coelliptical configurations are not permitted.  Approximate formula for small ellipticities are derived. These results reinforce and prolongate known results and classical theorems restricted to small elliptiticities. Comparisons with the numerical solutions computed from the Self-Consistent-Field method are successful.
\end{abstract}

\begin{keywords}
Gravitation | stars: interiors | stars: rotation | Methods: analytical
\end{keywords}

\section{Introduction} 

The theory of figures, indisputably, is among the most important production in theoretical astrophysics that has occupied mathematicians and physicists for more than two centuries \citep{chandra69,hachisu86III,horedttextbook2004,tohlinewiki21}. Since the pionnering contributions by Newton, Maclaurin, Jacobi and others, it has undergone various extensions and remains a reference to understand the structure of stars, planets, asteroids and even galaxies. In these contexts, the question of stratification in mass and in rotation is fundamental \citep[e.g.][]{sc42,maeder71,rcc15,cis19}. The equilibrium of a heterogeneous body made of homogeneous layers bounded by pure spheroidal surfaces has been soon investigated, with a special interest for the Earth and planets \citep{poincare88,alma990001099760306161}. According to Poincar\'e's theorem, only confocal configurations are compatible with solid rotation and lead to exact solutions. These states, however, require a mass-density inversion and seem therefore of minor interest \citep{hamy90,mmc83}. As quoted by \cite{hamy89}, approximate solutions are compatible with rigid rotation for small ellipticities, which situation is traditionnally reached in the slow-rotation limit. \cite{veronet12} has discussed the possibility of asynchronously rotating layers; see also \cite{mmc83}. In the present article, we reconsider Hamy's idea by expanding the gravitational potential over {\it confocal parameters} (instead of ellipticities), and include the hypothesis that layers are eventually in relative motion. As the degrees of freedom are increased, a larger diversity of equilibria follows. The approach is purely analytical. Numerical solutions remains technically tricky to obtain when mass-density jumps and rotational discontinuities are both present \citep{kiu10,ka16,bh21}. Actually, the surfaces bounding layers more or less deviate from pure spheroids and do not match with the standard coordinate systems. In addition, for classic computational grids, Poisson-solvers have generally poor efficiency in the presence of sharp density profiles.

We generalize the approach presented in \cite{h21a} (hereafter, Paper I) devoted to the $2$-layer problem, by considering ${\cal L}$ homogeneous layers in relative orbital motion. We mainly determine the rotation rates of each layer, the interface pressures and the central pressure as a function of the mass densities in the layers and the geometry of the spheroids (ellipticities and fractional radii). As for the two-layer case, the formalism resides on the expansion of the gravitational potential in the confocal parameters. Two classes of solutions can be distinguished, depending on the run of the gas pressure (constant or variable) along the interfaces: the ones associated with global rotation and the others corresponding to asynchronously rotating layers, respectively. The pertinent equations of problem, and in particular the conditions required for approximate rigid rotations, are presented in Sect. \ref{sec:equations}. The sequence of rotation rates $\{\Omega_1,\dots,\Omega_{\cal L}\}$, which is obtained by recursivity, is established in Sect. \ref{sec:solutions}. The problem is then recast in a compact, vectorial form. We briefly discuss the conditions for obtaining real rotation rates. Section \ref{sec:tests} is devoted to examples. Two configurations of special interest, namely the confocal and coelliptical states, are analyzed. The solutions obtained numerically from the Self-Consistent-Field (SCF) method with the multi-layer {\tt DROP}-code \citep{bh21} are used in comparison. We show in Sect. \ref{sec:typec} how to treat the special case of global rotation (all layers share the same rotation rate). In Sect. \ref{sec:lowrotation}, we derive the zero-order approximations for the $\Omega_i$'s valid for small ellipticities. The interface pressures along the rotation axis are given in the Appendix \ref{sec:pressures}. The two-layer case is reproduced in the Appendix \ref{sec:twolayer}. A basic Fortran 90 program that computes the sequence of rotation rates and the polar pressure at the interfaces is available in the Appendix \ref{sec:F90}. A summary and a few perspectives are found in the last section.

\begin{figure}
\includegraphics[width=8.4cm,bb=388 290 1067 791,clip==]{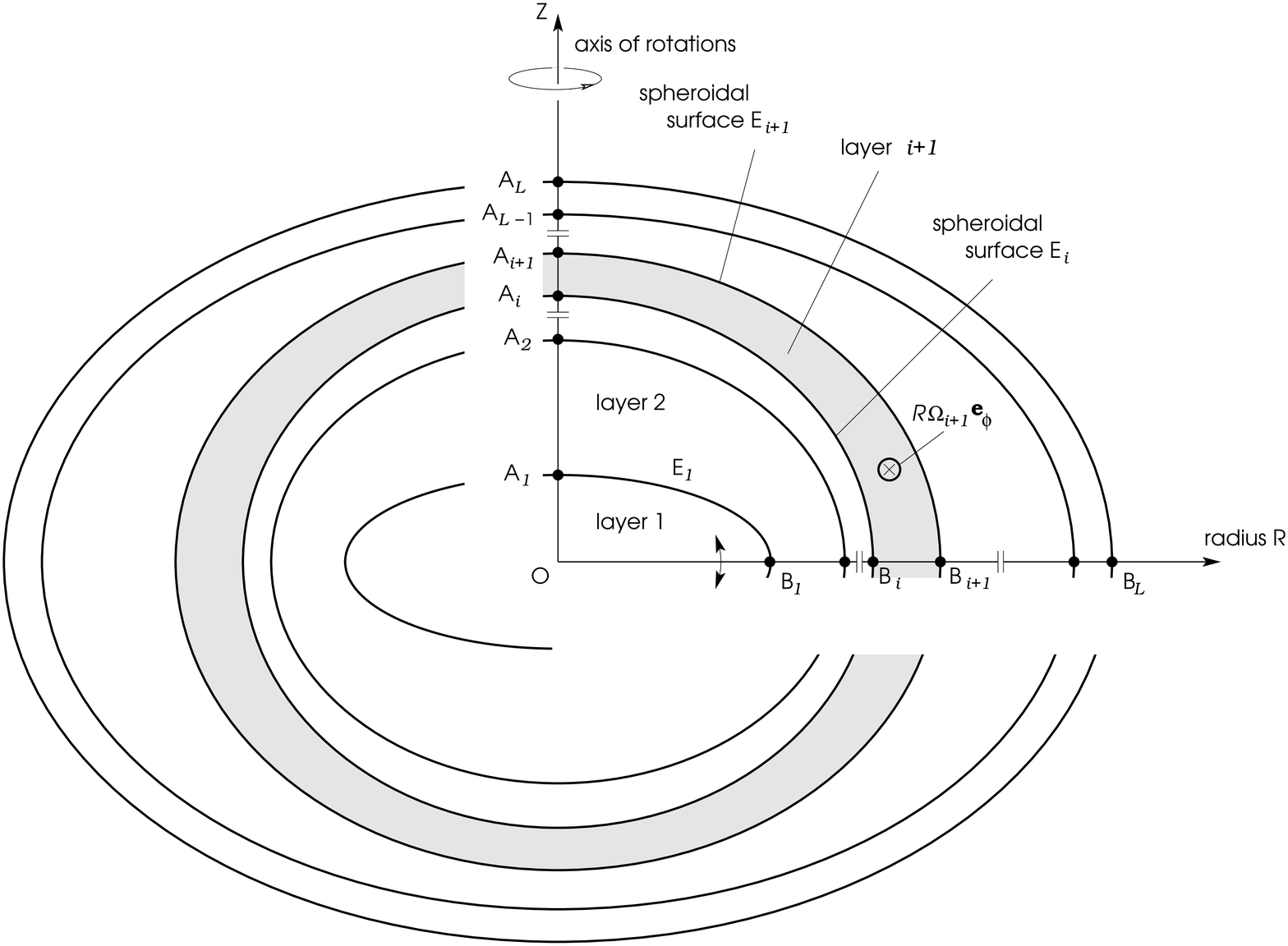}
\caption{Typical configuration for a nested structure made of ${\cal L}$ homogeneous layers (the deepest one is layer number $1$) bounded by oblate spheroidal surfaces. At each interface $E_i$, a mass-density jump and a rotational discontinuity are present; see (\ref{eq:defarhooa}) and Sect. \ref{sec:rigidrotations}.}
\label{fig:llayers.eps}
\end{figure}

\section{The equations of equilibrium}
\label{sec:equations}

\subsection{Theoretical background. Notations}

We adopt the same theoretical background and the same notations as in Paper I. We consider ${\cal L}$ oblate spheroidal surfaces $E_i(a_i,b_i)$ with semi-minor axis $b_i$ and semi-major axis $a_i>b_i$, with $i \in [1,{\cal L}]$, sharing the same axis of revolution and same plane of symmetry (and subsequently the same centre). Each surface $E_i$ (except the outermost one $E_{\cal L}$) is fully contained into a larger one $E_{i+1}$, i.e. $b_i<b_{i+1}$ and $a_i<a_{i+1}$, producing a nested structure as depicted in Fig. \ref{fig:llayers.eps}. The deepest one, bounded by $E_1$, has index $1$, while the outermost layer, bounded by $E_{{\cal L}-1}$ and $E_{\cal L}$, has index ${\cal L}$. Intersecting surfaces are excluded \citep[see e.g.][]{ca16}. If we introduce the fractional radius $\qij=a_i/a_j$, the immersion conditions write
\begin{subnumcases}{}
\barell_{i+1}- q_{i,i+1} \barell_i \ge 0, \label{eq:immersiona}\\
   q_{i,i+1} \le 1, \qquad i=1,\dots,{\cal L}-1 \label{eq:immersionb}
\end{subnumcases}
where $b_i/a_i=\barell_i$ is the dimensionless polar radius (or axis ratio) of $E_i$. While some constraints will be set in the sequel, there is no hypothesis yet on the ellipticities
\begin{flalign}
  \epsi = \sqrt{1-\barell_i^2},
\end{flalign}
which are allowed to differ from one surface to the other. These ${\cal L}$ surfaces define ${\cal L}$ domains (or layers), which are all homogeneous. If $\rho_i$ denotes the mass density of layer number $i$, the mass fractions $\nu_i$ are given by\footnote{As suggested by the referee, an extra layer (with index $0$) null size and null mass can be placed at the center of the coordinates. This would enable to merge the two formula in (\ref{eq:fmassothers}) into a single expression provided $q_{0,{\cal L}}=0$. Then, we would have $\lambda \equiv R^2 + Z^2$ and $f=0$ in (\ref{eq:psisinglei}), and the sum in (\ref{eq:psitot}) could start at $i=1$. \label{note:layer0}}
\begin{flalign}
  M\nu_{i} = \frac{4}{3}\pi a_{\cal L}^3
  \begin{cases}
   \rho_1 \bareone q_{1 {\cal L}}^3, \quad i=1,\\
   \rho_{i} \left( q_{i, {\cal L}}^3 \bar{\epsilon}_{i} -q_{i-1, {\cal L}}^3 \bar{\epsilon}_{i-1}\right), \; i \in [2, {\cal L}],
  \end{cases}
  \label{eq:fmassothers}
\end{flalign}
where $M$ is the total mass. In these conditions, the equilibrium of layer $i \in [1,{\cal L}]$ is governed by the Bernoulli equation \citep{lyttleton1953stability}
\begin{flalign}
  \frac{p_i}{\rho_i} +\Phi_i + \Psi = \const_i,
  \label{eq:bernoulli}
\end{flalign}
where $p_i$ is the pressure of matter, $\Phi_i=-\int{\Omega_i^2(R)RdR}$ is the centrifugal potential, $R$ is the cylindrical radius, $\Omega_i(R)$ is the rotation rate, and $\Psi$ is the total gravitational potential. For practical reasons (the centrifugal force vanishes on the rotation axis), the constant in the right-hand-side is preferentially determined at $R=0$. The other decisive equations come from the requirement of pressure balance at the connection between any pair of adjacent layers, namely
\begin{flalign}
  p_{i+1}|_{E_i}=p_i|_{E_i}, \qquad i \in [1,{\cal L}-1].
  \label{eq:pbalance}
\end{flalign}
For the outermost surface, we have $p_{\cal L}|_{E_{\cal L}}=0$ in the absence of any ambient pressure $p_a$ (we take $p_a=0$ in the paper througout). In fact, this latter surface condition can be incorporated into (\ref{eq:pbalance}) if the ambient medium is regarded as a supplementary layer, with number ${\cal L}+1$ and null pressure $p_a=p_{{\cal L}+1}=0$. Finally, the Poisson equation, which yields $\Psi$ from the mass-density field, is the last equation (see below). There are $2{\cal L}+1$ equations in total.

\subsection{The total gravitational potential}
\label{subsec:pot}

The potential in (\ref{eq:bernoulli}) is the major source of complexity as it is requires, in general, to solve the Poisson equation \citep{clement74,hachisu86III}. However, the fact that all bounding surfaces are, in any meridional plane, perfect ellipses brings a significant simplification since a closed form exceptionnally exists in such a case. Actually, let us remind that the potential in space due to a homogeneous body (with mass density $\rho$) bounded by a spheroidal surface $E(a,b)$ can be written in the following compact form \citep[][]{chandra69,binneytremaine87}
\begin{flalign}
  \label{eq:psisinglei}
 & \frac{\Psi(R,Z)}{-\pi G \rho} = f\left[ A_0(\epsilon')(a^2+\lambda)  -A_1(\epsilon')R^2 -A_3(\epsilon')Z^2 \right],
\end{flalign}
where
\begin{subnumcases}{}
    A_0(\epsilon)=2\frac{\bar{\epsilon}}{\epsilon}\arcsin \epsilon,\label{eq:IA1A3a}\\
    A_1(\epsilon)=\frac{\bar{\epsilon}}{\epsilon^3}\left[\arcsin\epsilon-\epsilon\bar{\epsilon}\right],\label{eq:IA1A3b}\\
    A_3(\epsilon)=-2\frac{\bar{\epsilon}}{\epsilon^3}\left[\arcsin\epsilon-\frac{\epsilon}{\bar{\epsilon}}\right],\label{eq:IA1A3c}\\
    f=\frac{a^2b}{(a^2+\lambda)\sqrt{b^2+\lambda}},\label{eq:IA1A3d}\\
    {\epsilon'}^2 = 1 - \frac{b^2+\lambda}{a^2+\lambda},\label{eq:IA1A3e}
\end{subnumcases}
and $\lambda$ is defined by
\begin{subnumcases}{}
   \lambda=0, \quad  \text{inside the body and onto } E, \label{eq:lambdaa}\\
   \frac{R^2}{a^2+\lambda}+  \frac{Z^2}{b^2+\lambda} -1 =0, \label{eq:lambdab} \\
   \qquad \qquad \qquad \qquad \text{outside the body}.\nonumber
\end{subnumcases}
This formula therefore works indifferently inside and outside the body.  In (\ref{eq:lambdab}), $\lambda$ is the largest root of the second degree polynomial. Although not explicitely quoted, this quantity (and subsequently $f$ and $\epsilon'$) depend on $R$ and $Z$ outside $E$, with $a$ and $b$ as parameters. On this basis, the total potential of the nested structure is easily derived from the superposition principle by using (\ref{eq:psisinglei}) $2 {\cal L}-1$ times with appropriate settings for the parameters $a$ and $b$ of the ellipses in the sample, as each layer (except the deepest one) is bounded by two surfaces. We therefore have (see note \ref{note:layer0})
\begin{flalign}
  \label{eq:psitot}
  & \frac{\Psi(R,Z)}{-\pi G}= \sum_{i=2}^{\cal L} \rho_i \left\{ f_i \left[ A_0(\epsilon_i')(a_i^2+\lambda_i) -A_1(\epsilon_i')R^2 \right.\right.\\
  \nonumber
  &\left. -A_3(\epsilon_i')Z^2 \right] - f_{i-1} \left[ A_0(\epsilon_{i-1}')(a_{i-1}^2+\lambda_{i-1}) -A_1(\epsilon_{i-1}')R^2\right. \\
  \nonumber
  & \left.\left. -A_3(\epsilon_{i-1}')Z^2 \right] \right\}+ \rho_1 f_1 \left[ A_0(\epsilon_{1}') (a_1^2+\lambda_1) -A_1(\epsilon_1')R^2 \right.\\
  \nonumber
  & \qquad \left.  -A_3(\epsilon_1')Z^2 \right],
\end{flalign}
where $\lambda_i$, $f_i$ and $\epsilon_i'$ are still defined according to (\ref{eq:lambdaa})-(\ref{eq:lambdab}), (\ref{eq:IA1A3d}) and (\ref{eq:IA1A3e}) respectively, but $(a,b)$ must be replaced by $(a_i,b_i)$.

Given the objectives of the article, the potential is not needed everywhere in space, but only along each spheroid $E_i$ in the sample. More precisely, we have to consider pairs $(E_i,E_j)$ of surfaces. Let $\lambdaij$ designate the values of $\lambda_i$ associated with the spheroid $i$ bounded by $E_i(a_i,b_i)$ taken along $E_j(a_j,b_j)$, i.e. $\lambdaij=\lambda_i(E_j)$. In a similar way, $\fij$ and $\epsijprim$ stand respectively for $f_i$, $\epsilon_i$ evaluated onto $E_j(a_j,b_j)$. As $R$ and $Z$ along $E_j$ are linked by
\begin{flalign}
  Z^2=b_j^2(1-\varpi_j^2),
  \label{eq:zontoej}
\end{flalign}
where $\varpi_j=R/a_j$, we see that $\lambdaij$ depends on a single space variable, for instance the radius, i.e. $\lambdaij \equiv \lambdaij(\varpi_j)$ which, in turn, depends on $3$ parameters $b_j$, $a_i$ and $b_i$.

\subsection{The confocal parameters}

Although we will leave (\ref{eq:psitot}) in its actual form, the summation can be split into two terms: one term corresponding to  the potential of all layers located above a given layer $j$ in the sample (including the layer itself), and one term for the total potential of all layers located below it. This separation is explicit in the dissertation by \cite{hamy89}. Actually, we see from (\ref{eq:IA1A3d}) and (\ref{eq:IA1A3e}) that, if $E_j \subset E_i$, which occurs for $j \le i$, then $\fij=1$ and $\epsijprim=\epsilon_i$ since $\lambdaij=0$. On the contrary, if $E_j$ is exterior to $E_i$, which occurs for $j > i$, then $\lambdaij$ must be calculated from (\ref{eq:lambdab}) for any term of the summation. This second situation is the main source of difficulty.

For convenience, the quadratic equation in $\lambdaij$ is rewritten in terms of the new variable $\xij$ defined by
\begin{flalign}
  \xij a_j^2=a_i^2+\lambdaij,
  \label{eq:lambda2x}
\end{flalign}
and so, according to (\ref{eq:lambdab}), the relevant root is given by
\begin{flalign}
  \label{eq:xijgen}
  \xij&=\frac{1}{2}\left(1+\cij+\epsilon_j^2 \varpi_j^2\right)\\
  \nonumber
  &\qquad +\frac{1}{2}\sqrt{(1+\cij+\epsilon_j^2 \varpi_j^2)^2-4 \qij ^2 \epsilon_i^2 \varpi_j^2},
\end{flalign}
where
\begin{flalign}
  c_{i,j}=q_{ij}^2 \epsilon_i^2 - \epsilon_j^2
  \label{eq:confocalc}
\end{flalign}
is the {\it confocal parameter} associated with the pair $(E_i,E_j)$. There are ${\cal L}^2$ parameters of this kind in total. Note that any surface is confocal with itself, i.e. $c_{i,i}=0$. We see that, if $E_i$ is confocal with $E_j$, then $\xij=1$ since $\cij=0$, which just means that $\lambdaij$ is a constant. Otherwise, $\xij$ varies along $E_j$. We have $\xij=1+\cij$ at point A$_j$ of the polar axis and $\xij=1$ at point B$_j$ of the equatorial plane (see Fig. \ref{fig:llayers.eps}). By rearranging the summation in (\ref{eq:psitot}), the total potential at a radius $R= \varpi_j a_j$ onto $E_j$, denoted $\Psi(E_j,\varpi_j)$ from now on, can be rewritten as
\begin{flalign}
  \label{eq:psitotbis}
  & \frac{\Psi(E_j,\varpi_j)}{-\pi G a_{\cal L}^2 q_{j, {\cal L}}^2}= \sum_{i=1}^{{\cal L}-1} \left(\rho_i-\rho_{i+1}\right) \fij \left[ A_0(\epsijprim)\xij  \right.\\
  \nonumber
  &\qquad \qquad  \qquad  \qquad \left.-A_1(\epsijprim) \varpi_j^2  -A_3(\epsijprim)\barej^2(1-\varpi_j^2) \right]\\
  \nonumber
  &+\rho_{\cal L} \left[ A_0(\epsilon_{{\cal L}, j}')x_{{\cal L} j} -A_1(\epsilon_{{\cal L}, j}') \varpi_j^2  -A_3(\epsilon_{{\cal L}, j}')  \barej^2(1-\varpi_j^2) \right],
\end{flalign}
where we have used (\ref{eq:zontoej}), and
\begin{subnumcases}{}
    \xij = \qij^2 \label{eq:Asbisa}\\
    \fij=1,\label{eq:Asbisb}\\
    \epsijprim=\epsilon_i \quad  \text{if } E_j \subseteq E_i \; (\text{or } j \le i),\label{eq:Asbisc}\\
    \quad \text{or } \nonumber\\
    \xij  \text{ from } (\ref{eq:xijgen}) \text{ | see below},\label{eq:Asbisd}\\
        \fij=\frac{\qij^3 \barei}{\xij\sqrt{\xij-\qij^2 \epsi^2}},\label{eq:Asbise}\\
    \epsijprim=\frac{\qij \epsi}{\sqrt{\xij}} \quad  \text{if } E_i \subset E_j \; (\text{or } j > i).\label{eq:Asbisf}
\end{subnumcases}
Note that the last term in the right-hand-side of (\ref{eq:psitotbis}) can be simplified since $E_{\cal L}$ is always external to all other surfaces, meaning $\epsilon_{{\cal L}, j}'=\epsilon_{\cal L}$ for any $j \in [1,{\cal L}]$.

\begin{figure*}
  \includegraphics[height=17.5cm,bb=60 90 240 730,clip==, angle=-90]{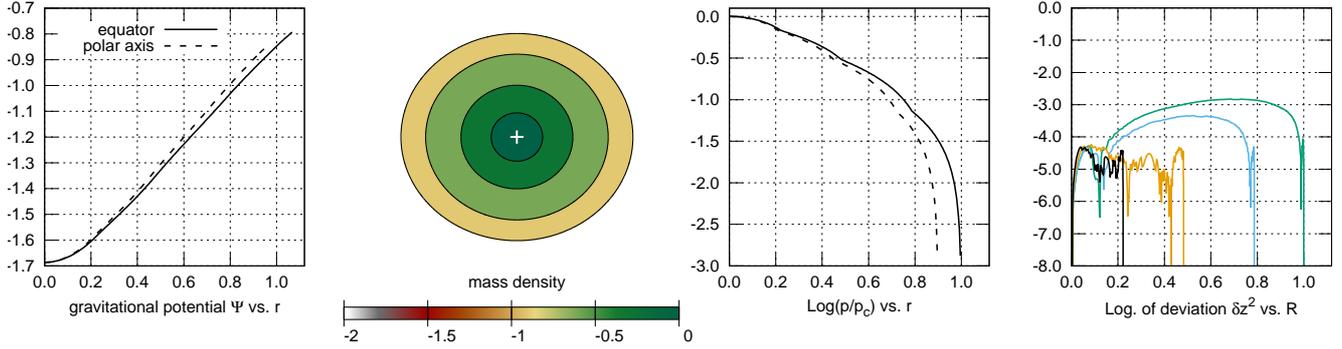}
\caption{Equilibrium solution computed for ${\cal L}=4$ with the {\tt DROP}-code for configuration A (see Tab. \ref{tab:resultsA} for the input parameter set and results). All layers rotate at the same rate (global rotation). {\it From left to right} (in log. scale): the gravitational potential at the pole  ({\it dashed line}) and at the equator ({\it plain line}), the normalized mass-density $\rho/\rhoc$ in color code, the normalized pressure $p/\pc$ and absolute deviations between the ``true'' surfaces and the ellipses $E_i$.}
\label{fig:graphrhoA.ps}
\end{figure*}

\begin{table}
  \centering
  \begin{tabular}{lrrrr}
    \multicolumn{5}{c}{{\bf configuration A} (input)}\\
   layer & $i=1$ & $2$ & $3$ & $4$ \\ \hline \hline
    $\leftarrow  q_{i,4}\bar{\epsilon}_i$  & $0.21$    &  $0.45$   & $0.72$   &  $0.9$  \\
  $\leftarrow \rho_i/\rho_{i+1}$                                   & $2$ & $2$ & $2$ \\ \hline \hline
  \end{tabular}\\\bigskip
  \begin{tabular}{lrrr}
    & {\tt DROP}-code & this work\\\hline
    $q_{1,4}$ & \multicolumn{2}{c}{$0.22176$} \\
    $q_{2,4}$ & \multicolumn{2}{c}{$0.48230$} \\
    $q_{3,4}$ & \multicolumn{2}{c}{$0.78690$} \\
    $\epsilon_1$ & \multicolumn{2}{c}{$0.32141$} \\
    $\epsilon_2$ & \multicolumn{2}{c}{$0.35984$} \\
    $\epsilon_3$ & \multicolumn{2}{c}{$0.40349$} \\ 
    $c_{1,4}$ & \multicolumn{2}{c}{$-0.18491$} \\
    $c_{2,4}$ & \multicolumn{2}{c}{$-0.15987$} \\
    $c_{3,4}$& \multicolumn{2}{c}{$-0.08918$}  \\ 
    $V/a_{\cal L}^3$ & $3.76482$ & $3.76991$ \\
    $\pc/\pi G \rho_{4} a_{4}^2$    & $6.22120$ & $6.22023$\\ 
    $p_1^*/\pi G \rho_{4} a_{4}^2$  & $4.23407$ &  $4.23393$ \\
    $p_2^*/\pi G \rho_{4} a_{4}^2$  & $1.89163$ &  $1.89159$\\
    $p_3^*/\pi G \rho_{4} a_{4}^2$  &  $0.45253$ & $0.45271$\\
    $\tilde{\Omega}_1^2$ & $0.13002$ & $0.13347$ & $0.13446^a$\\
    $\tilde{\Omega}_2^2$ & $0.13002$ & $0.13046$ & $0.12795^a$\\
    $\tilde{\Omega}_3^2$ & $0.13002$ & $0.13020$ & $0.12514^a$\\
    $\tilde{\Omega}_4^2$ & $0.13002$ & $0.13016$ & $0.12389^a$\\
    $M/\rho_4 a_{4}^3$ & $6.68126$ & $6.68742$ \\
    $\nu_1$ & $0.05178$ & $0.05175$ \\
    $\nu_2$ & $0.23664$ & $0.23639$ \\
    $\nu_3$ & $0.42743$ & $0.42737$ \\ \hline
    \multicolumn{4}{l}{$^*$value on the polar axis}\\
    \multicolumn{4}{l}{$^a$from (\ref{eq:omegalapprox}) and (\ref{eq:omegajapprox})}\\
  \end{tabular} 
   \caption{Input and output data for configuration A (${\cal L}=4$) associated with Fig. \ref{fig:graphrhoA.ps} obtained with the {\tt DROP}-code (column 2; see text) and from the actual formalism (columns 3 and 4); see notes \ref{note:units} and \ref{note:drop}; see also Sect. \ref{sec:tests}. Numbers are truncated.}
  \label{tab:resultsA}
\end{table}

\subsection{The conditions for approximate rigid rotations}
\label{sec:rigidrotations}

Since (\ref{eq:xijgen}) is not quadratic in $R$ and $\epsijprim$ are not constants, the centrifugal potentials $\Phi_i$ in (\ref{eq:bernoulli}) cannot, in general, be pure functions of $R^2$, which would correspond to strict, rigid rotations. Thus, we see that there is no exact solution to the problem of nested figures of equilibrium for rigid rotations if the bounding surfaces are perfect ellipsoids of revolution. This is not a new result. The exception is for configurations where $\cij=0$ for all pairs of surfaces, the so-called confocal configurations \citep{poincare88}. In such a case, because the $\lambdaij$'s, and subsequently the $f_i$'s and the $\epsilon_i'$'s are all constants, the ``coefficients'' $A_0$, $A_1$ and $A_3$  in (\ref{eq:IA1A3a})-(\ref{eq:IA1A3c}) do not depend on $R$ and $Z$, and are ``true'' constants. As pointed out in Paper I, there is another possibility: the presence of an ambient pressure acting at $E_{\cal L}$ can help to produce a series of rigid rotations in the system. This pressure has to absorb or compensate all terms in the potential at $E_{\cal L}$ of degree equal (optional) or higher than $R^2$. If some residuals remain, the compensation must continue down to the innermost surface $E_1$. As quoted above, we work with $p_a=0$.

As considered soon by \cite{hamy89}, approximate solutions compatible with rigid rotation can be derived in the case of small ellipticities, which correponds to slow rotations. By rigid rotation, we mean that each layer has its own solid body motion (implying a rotational discontinuity $\Omega_i/\Omega_{i+1}$ at each surface $E_i$). As in Paper I, we go beyond this hypothesis by assuming
\begin{flalign}
  |\cij| \ll 1 \qquad (i,j) \in [1,{\cal L}]^2,
  \label{eq:confcond}
\end{flalign}
which is more general than the combined conditions $\epsilon_i^2 \ll 1$ and $\epsilon_j^2 \ll 1$. Under these circumstances, we have from (\ref{eq:xijgen})
\begin{flalign}
  \xij \approx 1 + \cij (1 - \varpi_j^2), 
  \label{eq:xij}
\end{flalign}
in the first-order (see Paper I). At order zero in $\cij$, we have $\xij = 1$, and the $\lambdaij$'s are all constants, as well as all quantities in (\ref{eq:IA1A3a})-(\ref{eq:IA1A3e}). At order $1$, $\xij$ is quadratic with the radius and we have in particular
\begin{flalign}
  \begin{cases}
   \xij|_{{\rm A}_j}=1+\cij,\\
   \xij|_{{\rm B}_j}=1.
 \label{eq:xab}
 \end{cases}
\end{flalign}
at the two end-points A$_j$ and B$_j$ of $E_j$ (see Fig. \ref{fig:llayers.eps}; see again Paper I for more details).

We show in Fig. \ref{fig:graphrhoA.ps} an example\footnote{In the graphs and tables, the pressure is given in units of $\pi G\rhol^2 a_{\cal L}^2$, the rotation rates are given in units of $\sqrt{2 \pi G \rho_{\cal L}}$ and the potential is in units of $\pi G \rhol a_{\cal L}^2$ \label{note:units}.} of a nested figure of equilibrium obtained with $4$ layers (i.e. ${\cal L}=4$) with the {\tt DROP}-code that solves the full problem\footnote{In the {\tt DROP}-code, the interfaces at equilibrium are automatically detected all along the convergence cycle, which guarantees a good accuracy of various integrals involved (i.e., boundary conditions and global, output quantities). In its current version, the location of points B$_i$ in the equatorial plane is not known in advance. Like the ellipticities and fractional radii, these are outputs.
  In practice, we have adopted resolution of $1/256^2$, which is reached for $8$ levels of multigrid. Numbers in the tables are therefore limited to $5$ digits, and truncated. \label{note:drop}} by numerical means from the Self-Consistent-Field (SCF) method \citep{bh21}. The rigid rotation law is used, and all layers rotate at the same rate here (rotation is global). Table \ref{tab:resultsA} contains all input parameters (ellipticities, mass-density jumps and fractional sizes) and the main output quantities (interface pressure, rotation rate, fractional masses, etc.; see note \ref{note:units}). Unsurprisingly, the interfaces are not perfect ellipses, mainly because the true gravitational potential is not strickly given by (\ref{eq:psitot}). The deviations shown in Fig. \ref{fig:graphrhoA.ps}, in $Z^2$, are, however, less than $10^{-3}$ in absolute for all the surfaces involved. The confocal parameters are negative, less than $0.19$ in absolute. These results indicate that the hypothesis of nested figures of equilibrium based on spheroidal surfaces and approximate rigid rotations is fully justified (more examples below).

\section{Solutions}
\label{sec:solutions}

As long as the approximation of rigid rotation holds, the solution to the problem is obtained from (\ref{eq:bernoulli}), (\ref{eq:pbalance}) and (\ref{eq:psitot}). It is fully analytical. As in Paper I, we can determine the interface pressures at $R=0$ where the centrifugal force vanishes. This is summarized in the Appendix \ref{sec:pressures}.
      
\subsection{The sequence of rotation rates}

The rotation rate of the top layer is obtained first, because $E_{\cal L}$ is a surface of null pressure. From (\ref{eq:bernoulli}) evaluated at the two end-points $A_{\cal L}$ and $B_{\cal L}$, where $\varpi_{ij}=0$ and $1$  respectively, and by using (\ref{eq:psitotbis}) for $j={\cal L}$, we find after substraction
\begin{flalign}
- \frac{1}{2}\Omega_{\cal L}^2 a_{\cal L}^2 + \Psi(E_{\cal L},1)- \Psi(E_{\cal L},0)= 0,
  \label{eq:bernoullitop}
\end{flalign}
which leads to
\begin{flalign}
  \label{eq:omegatop}
 &\frac{\Omega_{\cal L}^2}{2 \pi G} =  \rhol \left[ A_1(\epsilon_{\cal L}) - A_3(\epsilon_{\cal L}) \bar{\epsilon}_{\cal L}^2 \right]\\
  \nonumber
 & - \sum_{i=1}^{{\cal L}-1} (\rho_i-\rho_{i+1}) \left.  \left\{ f_{i {\cal L}} \left[ A_0(\epsijprim)x_{i {\cal L}}  -A_3(\epsijprim) \bar{\epsilon}_{\cal L}^2\right] \right\}\right|_{{\rm A}_{\cal L}}^{{\rm B}_{\cal L}}\\
  \nonumber
  & \qquad + \sum_{i=1}^{{\cal L}-1} (\rho_i-\rho_{i+1}) \left. \left\{ f_{i {\cal L}} \left[ A_1(\epsilon_{i {\cal L}}') - A_3(\epsilon_{i {\cal L}}') \bar{\epsilon}_{\cal L}^2 \right] \right\}\right|_{{\rm B}_{\cal L}}.
\end{flalign}
The first term inside the brackets in the right-hand-side is nothing but the Maclaurin function
\begin{flalign}
{\cal M}(\epsilon)=A_1(\epsilon)- (1-\epsilon^2) A_3(\epsilon),
\end{flalign}
corresponding to a single Maclaurin spheroid bounded by $E_{\cal L}$ (see Paper I). The other terms come from the embedded layers (with mass density in excess of $\rhol$). These can be expressed by using the intermediate functions ${\cal P}$ and ${\cal C}$ already defined in Paper I, namely
\begin{flalign}
  {\cal M}(\epsilon) {\cal P}(\epsilon,\epsilon') = A_3(\epsilon')(1-\epsilon^2)-A_1(\epsilon'),
 \label{eq:pfunction}
\end{flalign}
and
\begin{flalign}
\label{eq:dq}
      {\cal M}(\epsilon) {\cal C}(\epsilon,\epsilon')&= A_0(\epsilon')x - (1-\epsilon^2) A_3(\epsilon'),
\end{flalign}
where $x$ is linked to $\lambda$ according to (\ref{eq:lambda2x}). On this basis, and with the supplementary definitions
\begin{subnumcases}{}
    \alpha_i=\frac{\rho_i}{\rho_{i+1}},  \label{eq:defarhooa}\\
  \trho_i=\frac{\rho_i}{\rhol},  \label{eq:defarhoob}\\
  \tOmega_i^2=\frac{\Omega_i^2}{2\pi G \rhol},  \label{eq:defarhooc}
\end{subnumcases}
we see that (\ref{eq:omegatop}) becomes
\begin{flalign}
  \nonumber
  &\tilde{\Omega}_{\cal L}^2 = {\cal M}(\epsl) \left\{1- \sum_{i=1}^{{\cal L}-1} \trho_{i+1} (\alpha_i-1)   \left. f_{i {\cal L}}{\cal P}(\epsilon_{\cal L},\epsilon_{i, {\cal L}}') \right|_{B_{\cal L}} \right. \\
  &\left. \qquad \qquad - \sum_{i=1}^{{\cal L}-1} \trho_{i+1} (\alpha_i-1) \underbrace{ \left.  f_{i {\cal L}}  {\cal C}(\epsilon_{\cal L},\epsilon_{i, {\cal L}}')\right|_{A_{\cal L}}^{B_{\cal L}}}_{\text{corrections}} \right\},
  \label{eq:omegal}
\end{flalign}
where $\left. {\cal P}(\epsilon_{\cal L},\epsilon_{i, {\cal L}}') \right|_{B_{\cal L}} = {\cal P}(\epsilon_{\cal L},q_{i, {\cal L}}\epsilon_{i, {\cal L}})$. As evoked already, if $E_i$ and $E_{\cal L}$ are confocal, then $x_{i, {\cal L}}=1$ both at $A_{\cal L}$ and B$_{\cal L}$ for a given $i$, with the consequence that $\left. f_{i, {\cal L}} {\cal C}(\epsilon_{\cal L},\epsilon_{i, {\cal L}}')\right|_{A_{\cal L}}^{B_{\cal L}}=0$. The last summation therefore represents the series of $1$rst-order corrections with respect to confocal configurations. It is expected to be small if all the $\cij$'s are close to zero. Note that the amplitude of each corrective term with respect to the corresponding leading one does not depend on $\alpha_i$ but just on the geometry of the pair of ellipses involved.

Once $\tilde{\Omega}^2_{\cal L}$ is known, we can deduce the rotation rate of all layers down to the centre by recursion, as follows. If we multiply the Bernoulli equation for layer $j < {\cal L}$ by the mass-density jump $\alpha_j$ at the interface with layer $j+1$ where $p_j=p_{j+1}$ along $E_j$ by virtue of (\ref{eq:pbalance}), we get from (\ref{eq:bernoulli})
\begin{flalign}
  \label{eq:omegarecursion1}
 \frac{1}{2}\left(\Omega_{j+1}^2-\alpha_j \Omega_{j}^2\right)R^2& +(\alpha_j-1) \Psi(E_j,\varpi_j)\\
  \nonumber
 &\qquad  +\const_{j+1}-\alpha_j \const_j=0,
\end{flalign}
where $\Psi(E_j,\varpi_j)$ is given by (\ref{eq:psitotbis}). The two constants are easily eliminated by evaluating this expression at the two end-points of the surface $E_j$, namely at point A$_j$ where $\varpi_j=0$ and at point B$_j$ where $\varpi_j=1$ (see Fig. \ref{fig:llayers.eps}). After some algebra and simplication by $q^2_{j, {\cal L}} \ne 0$, we get
\begin{flalign}
  \label{eq:omegarecursion2}
  &\frac{\tilde{\Omega}_{j+1}^2 - \alpha_j \tilde{\Omega}_j^2}{\alpha_j-1}\\
  \nonumber
  &- \sum_{i=1}^{{\cal L}-1} \trho_{i+1}(\alpha_i-1) \left. \left\{ \fij \left[ A_0(\epsijprim) \xij -A_1(\epsijprim)\right] \right\} \right|_{{\rm B}_j}  \\
  \nonumber
  & \qquad + \sum_{i=1}^{{\cal L}-1}  \trho_{i+1} (\alpha_i-1) \left. \left\{ \fij \left[ A_0(\epsijprim) \xij -A_3(\epsijprim) \barej^2 \right] \right\} \right|_{{\rm A}_j}\\
  \nonumber
  &\qquad\qquad+ \left. f_{{\cal L}, j} \left[ A_0(\epsilon_{{\cal L}, j}') x _{{\cal L}, j} -A_1(\epsilon_{{\cal L}, j}')  \right] \right|_{{\rm B}_j}\\
  \nonumber
  & \qquad\qquad\qquad -  \left. f_{{\cal L}, j} \left[ A_0(\epsilon_{{\cal L}, j}') x _{{\cal L}, j} -A_3(\epsilon_{{\cal L}, j}') \barej^2 \right] \right|_{{\rm A}_j} =0
\end{flalign}
where $\alpha_j-1 \ne 0$ (otherwise $E_j$ is not an interface and layers $j$ and $j+1$ merge into a unique domain). The last two terms can be simplified because $j < {\cal L}$. Again, after substancial rearrangements and by using the functions ${\cal P}$ and ${\cal C}$, we find
\begin{flalign}
  \label{eq:omegarecursion3}
  &\alpha_j\tilde{\Omega}_j^2 = \tilde{\Omega}_{j+1}^2  - (\alpha_j-1) {\cal M} (\epsj) \\
  \nonumber
  &  \times \left\{ {\cal P}(\epsj,\epsilon_{\cal L}) + \sum_{i=1}^{{\cal L}-1}  \trho_{i+1} (\alpha_i-1) \left. \fij {\cal P}(\epsj,\epsijprim) \right|_{{\rm B}_j} \right.  \\
  \nonumber
  & \qquad + \left. \sum_{i=1}^{{\cal L}-1}\left. \trho_{i+1} (\alpha_i-1)  \fij {\cal C}(\epsj,\epsijprim)\right|_{A_{\cal L}}^{B_{\cal L}} \right\},
  \end{flalign}
which form clearly shows the leading term and the first-order correction. As ${\cal P}(\epsj,\epsj)=-1$, we see that this expression (\ref{eq:omegarecursion3}) even works for $j={\cal L}$ if we imagine that the whole nested structure is immersed into an hypothetical medium with extremely low density $\rho_{{\cal L}+1} \ll \rhol$ and rotating at the same rate as the outermost layer ${\cal L}$, i.e. $\tilde{\Omega}_{\cal L} = \tilde{\Omega}_{{\cal L} +1}$ (both sides of the equation then simplify by $\alpha_i-1 \ne 1$). This medium must not bring any contribution to the total gravity field. This is typically the situation of generalized Roche systems, where a massive body carries away a rarefied gas \citep[e.g.][]{jeans28,maeder2009}. In the present case, the rotation rate of this extra, massless component is the rotation rate of the outermost layer ${\cal L}$. 

\subsection{Vectorial notation}

It is clear that (\ref{eq:omegal}) and (\ref{eq:omegarecursion3}) can be regarded as scalar products in a ${\cal L}$-dimension Euclidean space (with, for instance, $\{\mathbf{u}_1,\dots,\mathbf{u}_{\cal L}\}$ as the natural basis of unit vectors). Actually, if we define the three vectors
\begin{flalign}
  \label{eq:xvect}
  \mathbf{X} =  \frac{1}{\rhol}
	\begin{pmatrix} 
		\rho_1-\rho_2 \\ 
		\rho_2-\rho_3 \\
                \dots\\
		\rho_i-\rho_{i+1}\\
                \dots\\
		\rho_{{\cal L}-1}-\rhol\\
                \rhol
	\end{pmatrix},
\end{flalign}
\begin{flalign}
  \label{eq:pvect}
   \mathbf{P}({\rm B}_j) = 
	\begin{pmatrix} 
		\left. f_{1,j}{\cal P}(\epsj,\epsilon_{1, j}')\right|_{{\rm B}_j} \\ 
		\left. f_{2,j}{\cal P}(\epsj,\epsilon_{2, j}')\right|_{{\rm B}_j} \\
                \dots\\
	       	\left. f_{i,j}{\cal P}(\epsj,\epsilon_{i, j}')\right|_{{\rm B}_j}\\
                \dots\\
	        \left. f_{{{\cal L}-1},j}{\cal P}(\epsj,\epsilon_{{{\cal L}-1}, j}')\right|_{{\rm B}_j}\\
                \left. {\cal P}(\epsj,\epsilon_{{\cal L}, j}')\right|_{{\rm B}_j}
	\end{pmatrix},
\end{flalign}
and
\begin{flalign}
  \label{eq:cvect}
   \mathbf{C}({\rm A}_j,{\rm B}_j) = 
	\begin{pmatrix} 
		\left. f_{1,j} {\cal C}(\epsj,\epsijprim)\right|_{{\rm A}_j}^{{\rm B}_j} \\ 
  	        \left. f_{2,j} {\cal C}(\epsj,\epsilon_{2, j}')\right|_{{\rm A}_j}^{{\rm B}_j} \\
                \dots\\
	       	\left. f_{i,j} {\cal C}(\epsj,\epsilon_{i, j}')\right|_{{\rm A}_j}^{{\rm B}_j}\\
                \dots\\
	        \left. f_{{{\cal L}-1},j}{\cal C}(\epsj,\epsilon_{{{\cal L}-1}, j}')\right|_{{\rm A}_j}^{{\rm B}_j}\\
                0
	\end{pmatrix},
\end{flalign}
then (\ref{eq:omegal}) becomes
\begin{flalign}
  &\tilde{\Omega}_{\cal L}^2 = - {\cal M}(\epsl) \, \mathbf{X} \cdot \left[ \mathbf{P}({\rm B}_{\cal L}) + \mathbf{C}({\rm A}_{\cal L},{\rm B}_{\cal L}) \right].
  \label{eq:omegalrecursionvector}
\end{flalign}
where we have used $\mathbf{P}({\rm B}_{\cal L}) \cdot \mathbf{u}_{\cal L} = P_{\cal L}({\rm B}_{\cal L})= \-1$. From (\ref{eq:omegarecursion3}), the recurrence relation, for $j \in [1,{\cal L}-1]$,  is
\begin{flalign}
  \label{eq:omegajrecursionvector}
  &\alpha_j\tilde{\Omega}_j^2 = \tilde{\Omega}_{j+1}^2 \\
  \nonumber
  & \qquad \qquad - (\alpha_j-1) {\cal M} (\epsj) \, \mathbf{X} \cdot \left[ \mathbf{P}({\rm B}_j) + \mathbf{C}({\rm A}_j,{\rm B}_j) \right].
  \end{flalign}
For ${\cal L}=2$, we recover the expressions reported in Paper I; see the Appendix \ref{sec:twolayer}.

\begin{table}
  \centering
  \begin{tabular}{lrrrr}
   layer  & $i=1$ & $2$ & $3$ & comment  \\ \hline \hline
   $q_{i,{\cal L}}$  & $0.6$ &  $0.7$  &  $1$ \\
    $\alpha_i$  & $1.5$ & $2$ &  \\ 
    $\; \rightarrow X_i$ & $1$ & $1$ & $1$ \\
   $\bar{\epsilon}_i$& $0.9$ & $0.8$ & $0.7$ & $\nabla \epsilon >0$\\ \hline \hline
  \end{tabular}\\\bigskip
  \begin{tabular}{lccc}
          & $j=1$ & $2$ & $3$  \\ \hline
    ${\cal M}(\epsilon_j)$ & $0.05202$  &$0.10067$ &$0.14451$\\
    ${\bf P}({\rm B}_j)$ & $\begin{pmatrix} $-1$\\$+0.64262$\\$+2.53737$\\\end{pmatrix}$ & $\begin{pmatrix} $-1.17636$\\$-1$\\$-0.14791$\\\end{pmatrix}$ & $\begin{pmatrix} $-0.90406$\\ $-1.15023$\\ $-1$\\\end{pmatrix}$\\
    $\tilde{\Omega}_j^2$ & $0.22481$ & $0.39393$ & $0.53219$ \\\hline
  \end{tabular}
  \caption{Values for ${\cal M}(\epsilon_j)$, ${\bf P}({\rm B}_j)$ and $\tilde{\Omega}_j^2$ obtained from (\ref{eq:xvect}), (\ref{eq:pvect}), (\ref{eq:omegalrecursionvector}) and (\ref{eq:omegajrecursionvector}) for a configuration with ${\cal L}=3$. This case corresponds to a positive gradient of ellipticity from the center to the surface of the body.}
   \label{tab:gradpos}
\end{table}

\subsection{Conditions of positivity}

As for ${\cal L}=2$ (see Paper I, and references therein), the positivity of all $\Omega_j^2$'s is not guaranteed at this level. Even if we assume that there is no density inversion from the centre to the surface, the components of vector ${\mathbf P}$ can be positive or negative depending on the relative geometry of the spheroids (see Fig. 4 in Paper I). This point is critical regarding the existence of an equilibrium structure and it must be examined in details for a given set of input parameters (ellipticities $\epsilon_i$, fractional sizes $\qij$, and mass-density jumps $\alpha_i$). In the case ${\cal L}>2$, it does not seem easy to make a precise inventory of allowed and forbidden equilibria, because of the high number of parameters involved, $3 {\cal L}-2$ in total. First, we expect $\lVert{\mathbf C} \rVert \ll \lVert{\mathbf P}\rVert$ provided confocal parameters are all small in absolute; see (\ref{eq:confcond}). Second, we see from (\ref{eq:omegalrecursionvector}) that $\Omega_{\cal L}^2 >0$ if the scalar product $\mathbf{X} \cdot \mathbf{P}({\rm B}_{\cal L})$ is negative. For stability reasons, states involving density inversions (i.e. $\alpha_i < 1$) are not desirable. The preference is therefore given to configurations with essentially negative components of $\mathbf{P}({\rm B}_{\cal L})$. By looking at the ${\cal P}(\epsilon,\epsilon')$-graph (again, see Fig. 4 in Paper I), we see that, at order $0$, negative values of ${\cal P}(\epsilon_{\cal L},\epsilon_{i, \cal L}')$ occur typically when $\epsilon_{\cal L} \gtrsim \epsilon_{i, \cal L}'$, i.e. for $c_{i, \cal L} \lesssim 0$. This condition is automatically fulfilled when $\epsilon_{\cal L} \gtrsim \epsilon_i \ge q_{i, {\cal L}} \epsilon_i$, which corresponds to an outermost spheroid more oblate than all other spheroids. We can use a similar qualitative argument to see what happens below layer ${\cal L}$. Actually, from the recursion formula (\ref{eq:omegajrecursionvector}), we see that $\Omega^2_j$ is {\it unconditionnally positive} if the ${\cal L}$ components $P_i({\rm B}_j)$ all take negative values. This occurs when $\epsilon_j \gtrsim \epsilon_{i,j}' \ge \qij \epsilon_i$, i.e. for $c_{i,j} \lesssim 0$ (again, see Fig. 4 in Paper I), namely when the spheroid $E_j$ is more oblate than those located below. The reasoning holds down to the core. Note that this does not depend on the mass-density jumps, provided $\alpha_i >1$. The most favorable and natural situations for the existence of nested figures are therefore expected for {\it increasing ellipticities from the center to the surface}, as in the two-layer case. This prolongates the result by \cite{hamy89}, which is limited to small ellipticities. If we omit the correction, (\ref{eq:omegajrecursionvector}) can be put in the form
\begin{flalign}
  \label{eq:gradomegaj}
  &\tilde{\Omega}_{j+1}^2-\tilde{\Omega}_j^2  \approx (\alpha_j-1)\left[\tilde{\Omega}_j^2+ {\cal M} (\epsj) \, \mathbf{X} \cdot \mathbf{P}({\rm B}_j)\right].
  \end{flalign}
As a consequence, when $\nabla \epsilon >0$, $\Omega_{\cal L}^2$ can be significantly large, while $\Omega_1^2$ is, in contrast, rather small because $P_1({\rm B}_j) \gtrsim 0$ for $j>1$. It follows that $\tilde{\Omega}_{j+1}^2-\tilde{\Omega}_j^2$ is expected to be positive, i.e. $\nabla \Omega > 0$. This is illustrated with the data of Tab. \ref{tab:gradpos}. Quite logically, the more oblate a layer the faster its rotation rate. This is however not a general rule. Configurations with negative gradients of ellipticity are also permitted. In such a case, $P_1({\rm B}_j) < 0$ typically. The second term in the right-hand side of (\ref{eq:gradomegaj}) can be large in absolute (but negative), leading to $\tilde{\Omega}_{j+1}^2-\tilde{\Omega}_j^2 < 0$. We give in Tab. \ref{tab:gradneg} an example of such a configuration where a negative gradient of ellipticity is associated with a negative gradient of rotation rate. Again, this is not a general rule, just a trend. It immediately follows from this discussion that if the configurations where the rotation rates of all layers are close to each other, i.e. $\nabla \Omega \approx 0$, then the $\epsilon_j$'s can increase or decrease from the center to the surface. As (\ref{eq:gradomegaj}) indicates, this is especially true when the $\alpha_i$'s are all close to unity. However, we do not have  $\nabla \epsilon \rightarrow 0$ in the limit $\nabla \Omega \rightarrow 0$, i.e. at global rotation (see below).

\begin{table}
  \centering
  \begin{tabular}{lrrrr}\\
   layer  & $i=1$ & $2$ & $3$ & comment  \\ \hline \hline
   $q_{i,{\cal L}}$  & $0.6$ &  $0.7$  &  $1$ \\
    $\alpha_i$  & $1.5$ & $2$ &  \\ 
    $X_i$ & $1$ & $1$ & $1$ \\
   $\bar{\epsilon}_i$& $0.7$ & $0.8$ & $0.9$ & $\nabla \epsilon <0$\\ \hline \hline
  \end{tabular}\\\bigskip
  \begin{tabular}{lccc}
          & $j=1$ & $2$ & $3$  \\ \hline
     ${\cal M}(\epsilon_j)$ & $0.14451$ & $0.10067$ & $0.05202$\\
   ${\bf P}({\rm B}_j)$ & $\begin{pmatrix} $-1$\\$-1.515512$\\$-1.96242$\\\end{pmatrix}$ & $\begin{pmatrix} $-0.51642$\\$-1$\\$-1.7387$\\\end{pmatrix}$ & $\begin{pmatrix} $-0.00780$\\ $-0.09029$\\ $-1$\\\end{pmatrix}$\\
          $\tilde{\Omega}_j^2$ & $0.34618$  & $0.19572$ & $0.06363$\\\hline
    \end{tabular}
 \caption{Same caption as for Tab. \ref{tab:gradpos} but for a negative gradient of ellipticity.}
  \label{tab:gradneg}
\end{table}

\section{Special cases and examples}
\label{sec:tests}

\subsection{Confocal configurations}
\label{sec:confocal}

\begin{figure}
\includegraphics[width=8.3cm,bb=85 65 504 417,clip==]{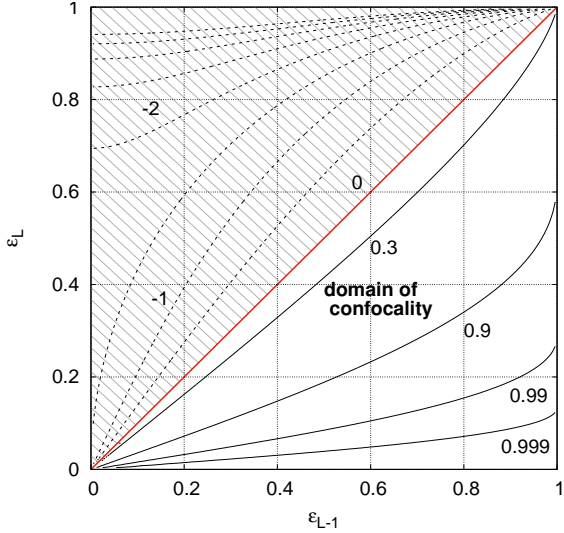}
\caption{The function $\mu(\epsilon_{{\cal L }-1},\epsilon_{\cal L })$ defined by (\ref{eq:mu}) in the form of levels of contours (values are labelled on the curves). In the domain relevant for confocality (bottom right-part of the graph), $\mu < 1$, which implies that confocal configurations are not compatible with layers ${\cal L}-1$ and ${\cal L}$ in synchronous rotation, for a normal/stable stratification of the mass density in the whole structure.}
\label{fig:confocality.eps}
\end{figure}

For confocal states, $\cij=0$ for all pairs $(E_i,E_j)$ of spheroidal surfaces, which means that $\lambdaij=\const$ along $E_i$. In these conditions, we have $\mathbf{C}({\rm A}_j, {\rm B}_j) = \mathbf{0}$, ${\cal P}(\epsilon_j,\epsijprim)=-1$ and $\fij=\qij^3\bar{\epsilon}_i/\bar{\epsilon}_j$ at point B$_j$. We easily show that, for $j={\cal L}$, (\ref{eq:omegalrecursionvector}) becomes
\begin{flalign}
   \label{eq:xpl}
 - \mathbf{X} \cdot \mathbf{P}({\rm B}_{\cal L}) = \frac{M}{\frac{4}{3}\pi \rhol a^3_{\cal L} \bar{\epsilon}_{\cal L}} \equiv \mu.
\end{flalign}
We see that the denominator in the right-hand-side is just the mass of a (single-component) Maclaurin spheroid having the same mass-density and same external bounding surface as the outer layer ${\cal L}$ of the nested structure. As a consequence, we have $\mu>1$ provided $\rhol < \rho_i$ for $i \in [1,{\cal L}-1]$, i.e. the outer layer is less dense than all the interior ones (no density inversion). Let us now determine the rotation rate of the layer located just below, by setting $j={\cal L}-1$ in the formula. We still have ${\cal P}(\epsi,\epsijprim)=-1$ for all components except for the last one, which is ${\cal P}(\epsilon_{\cal L},\epsilon_{{\cal L}-1, {\cal L}}')={\cal P}(\epsilon_{\cal L},\epsilon_{{\cal L}-1})$. The scalar product in (\ref{eq:omegajrecursionvector}) therefore writes
\begin{flalign}
 \nonumber
  &-\mathbf{X} \cdot \mathbf{P}({\rm B}_{{\cal L}-1}) = \frac{1}{\rhol}\left[(\rho_1-\rho_2)f_{1, {\cal L}-1}+(\rho_2-\rho_3)f_{2, {\cal L}-1}+ \dots \right.\\
  & \qquad  \qquad \qquad \qquad \left.+ (\rho_{{\cal L}-1}-\rhol) - \rhol {\cal P}(\epsilon_{\cal L},\epsilon_{{\cal L}-1}) \right],
  \label{eq:spl}
\end{flalign}
which can advantageously be rewritten as a function of the total mass, namely
\begin{flalign}
  \label{eq:splmoinsun}
  &-\mathbf{X} \cdot \mathbf{P}({\rm B}_{{\cal L}-1}) = \frac{1}{\rhol a^3_{{\cal L}-1} \bar{\epsilon}_{{\cal L}-1}}\left\{\frac{M}{\frac{4}{3}\pi} \right.\\
  & \qquad  \qquad \qquad \qquad \left.- \rhol \left[ a^3_{{\cal L}-1} \bar{\epsilon}_{{\cal L}-1}{\cal P}(\epsilon_{\cal L},\epsilon_{{\cal L}-1}) + a^3_{\cal L} \bar{\epsilon}_{\cal L}\right] \right\}.
 \nonumber
\end{flalign}
Let us now assume that layers ${\cal L}-1$ and ${\cal L}$ rotate at the same rate, i.e. $\Omega_{\cal L} = \Omega_{{\cal L}-1}$. From (\ref{eq:omegalrecursionvector}) and (\ref{eq:omegajrecursionvector}), and given the two scalar products (\ref{eq:spl}) and (\ref{eq:splmoinsun}), the condition of synchronous rotation implies
\begin{flalign}
  &f_{{{\cal L}-1}, {\cal L}}{\cal M}(\epsilon_{\cal L}) \mu \\
  &-  {\cal M}(\epsilon_{{\cal L}-1}) \left[ \mu - f_{{{\cal L}-1}, {\cal L}}{\cal P}(\epsilon_{\cal L},\epsilon_{{\cal L}-1}) - 1\right]=0,
  \nonumber
\end{flalign}
where
\begin{flalign}
  f_{{{\cal L}-1}, {\cal L}} = \frac{\epsilon^3_{\cal L} \bar{\epsilon}_{{\cal L}-1}}{\epsilon^3_{{\cal L}-1} \bar{\epsilon}_{\cal L}},
\end{flalign}
 by virtue of confocality. If we solve this equation for $\mu$, we get
\begin{flalign}
 \mu = \frac{{\cal M}(\epsilon_{{\cal L}-1}) \left[ 1 + f_{{{\cal L}-1}, {\cal L}} {\cal P}(\epsilon_{\cal L},\epsilon_{{\cal L}-1}) \right]}{{\cal M}(\epsilon_{{\cal L}-1}) - f_{{{\cal L}-1}, {\cal L}}{\cal M}(\epsilon_{\cal L})},
 \label{eq:mu}
\end{flalign}
and the question is: do we still have $\mu > 1$? Clearly, this new expression for $\mu$ depends on two parameters $\epsilon_{{\cal L }-1}$ and $\epsilon_{\cal L } < \epsilon_{{\cal L }-1}$. It is plotted in the form of contour levels in Fig. \ref{fig:confocality.eps}, and we find that $\mu$ is always less than unity. This is therefore in contradiction with the above conclusion; see (\ref{eq:xpl}). It means that layers ${\cal L}-1$ and ${\cal L}$ cannot rotate at the same rate. Note that no value has been assigned to ${\cal L}$ here. We therefore confirm that {\it a heterogeneous body made of homogeneous components separated by confocal spheroids cannot be in global rotation if the mass-density decreases from the center to the surface}, in the conditions of the actual approximation. This generalizes the result obtained by \cite{mmc83} for ${\cal L}=2$; see also Paper I; see Sect. \ref{sec:tests} for an illustration. Again, this agrees with \cite{hamy89}.

\subsection{Coelliptical configurations}
\label{sec:coell}

Configurations where all ellipsoidal surfaces have the same ellipticity is of special interest. It is for instance the main assumption made in the model of concentric Maclaurin spheroids (CMS) for planetary interiors \citep{hub13}. These states are obtained by setting $\epsilon_i=\epsilon$ for $i = [1,{\cal L}]$ in the equations. Let us consider the deepest layer, first. From (\ref{eq:omegajrecursionvector}) with $j=1$, we have
\begin{flalign}
  &\alpha_1\tilde{\Omega}_1^2 = \tilde{\Omega}_{2}^2  - (\alpha_1-1) {\cal M} (\epsilon) \, \mathbf{X} \cdot \left[ \mathbf{P}({\rm B}_1) + \mathbf{C}({\rm A}_1,{\rm B}_1) \right].
  \end{flalign}
There is no confocal correction for $j=1$ (whatever the set of ellipticities) since $E_1 \subset E_i$ for all surfaces in the sample. As $\epsilon_{i,1}'=\epsilon_i$ and $f_{i,1}=1$, we have ${\cal P}(\epsilon_1,\epsilon_{i, 1}')=-1$, and it follows that $\mathbf{X} \cdot \mathbf{P}({\rm B}_1) = \rho_1/\rhol$. The rotations rates of the deepest layers are therefore linked by the simple relationship
\begin{flalign}
  \label{eq:omega1recursionvector}
  &\alpha_1\tilde{\Omega}_1^2= \tilde{\Omega}_{2}^2 + (\alpha_1-1) {\cal M} (\epsilon) \frac{\rho_1}{\rhol}.
\end{flalign}
If we consider the surface $E_2$ which separates the layers $2$ and $3$ and set $j=2$ in (\ref{eq:omegajrecursionvector}), we get
\begin{flalign}
  \label{eq:omega2recursionvector}
  &\alpha_2\tilde{\Omega}_2^2 = \tilde{\Omega}_{3}^2 - (\alpha_2-1) {\cal M} (\epsilon) \, \mathbf{X} \cdot \left[ \mathbf{P}({\rm B}_2) + \mathbf{C}({\rm A}_2,{\rm B}_2) \right].
\end{flalign}
As above, $f_{i,2}=1$, $P_i(\epsilon_1,\epsilon_{i, 1}')=-1$ and $C_i({\rm A}_2,{\rm B}_2)=0$ for any $i>1$ (because layer $2$ is exterior to layer $1$). We then have
\begin{flalign}
  &-\rhol \, \mathbf{X} \cdot \left[ \mathbf{P}({\rm B}_2) + \mathbf{C}({\rm A}_2,{\rm B}_2) \right]=\\
  \nonumber
 & \qquad - (\rho_1-\rho_2) \left[ \left. f_{1,2} {\cal P}(\epsilon_2,\epsilon_{1,2}')\right|_{{\rm B}_2} +   \left. f_{1,2} {\cal C}(\epsilon_2,\epsilon_{1,2}')\right|_{{\rm A}_2}^{{\rm B}_2} \right] \\
  \nonumber
  & \qquad\qquad+(\rho_2-\rho_3)+(\rho_3-\rho_4)+\dots+(\rho_{{\cal L}-1}-\rho_{\cal L})+\rho_{\cal L},\\
 \nonumber
 & \quad\qquad\qquad\qquad\qquad\qquad\qquad= \rho_1 -(\rho_1-\rho_2) h,
\end{flalign}
where the function
\begin{flalign}
  \label{eq:lambdafunc}
  h = 1+\left. f_{1,2} {\cal P}(\epsilon_2,\epsilon_{1,2}')\right|_{{\rm B}_2} +   \left. f_{1,2} {\cal C}(\epsilon_2,\epsilon_{1,2}')\right|_{{\rm A}_2}^{{\rm B}_2},
\end{flalign}
has already been introduced in Paper I. Obviously, $\epsilon_2=\epsilon$ and $\epsilon_{1,2}'$ depends on $q$ and $\epsilon$, which makes $h$ a function of these two quantities. We see that (\ref{eq:omega2recursionvector}) then takes the form
\begin{flalign}
 \label{eq:omega2recursionvector_bis}
  &\alpha_2\tilde{\Omega}_2^2= \tilde{\Omega}_{3}^2 + (\alpha_2-1) {\cal M} (\epsilon) \frac{\rho_1}{\rhol}  \left[1 - \frac{\rho_1-\rho_2}{\rho_1}h(\epsilon,q) \right].
\end{flalign}
Let us assume that layers $1$ to $3$ are in synchronous rotation. We see from (\ref{eq:omega1recursionvector}) with (\ref{eq:omega2recursionvector_bis}) that this is possible only if $h(\epsilon,q\epsilon)=0$ somewehere in the range $(\epsilon,q) \in [0,1]^2$. As shown in Paper I (see Fig. 6), $h$ remains positive in the whole plane, and it vanishes only for $\epsilon=q\epsilon$ (namely $q=1$, i.e. layer $2$ has null size), which can easily be established from (\ref{eq:lambdafunc}). We conclude that layers $1$ to $3$ cannot rotate at the same rate (as shown in Paper I, synchronous rotation is not possible for ${\cal L}=2$). Note, again, that no value of ${\cal L}$ has been specified in the demonstration. We conclude that {\it a heterogeneous body made of homogeneous components separated by coelliptical/similar spheroids cannot be in global rotation}, in the conditions of the actual approximation. An example is given below. This agrees with Hamy's theorem.

\begin{figure*}
  \includegraphics[height=17.5cm,bb=60 90 240 730,clip==, angle=-90]{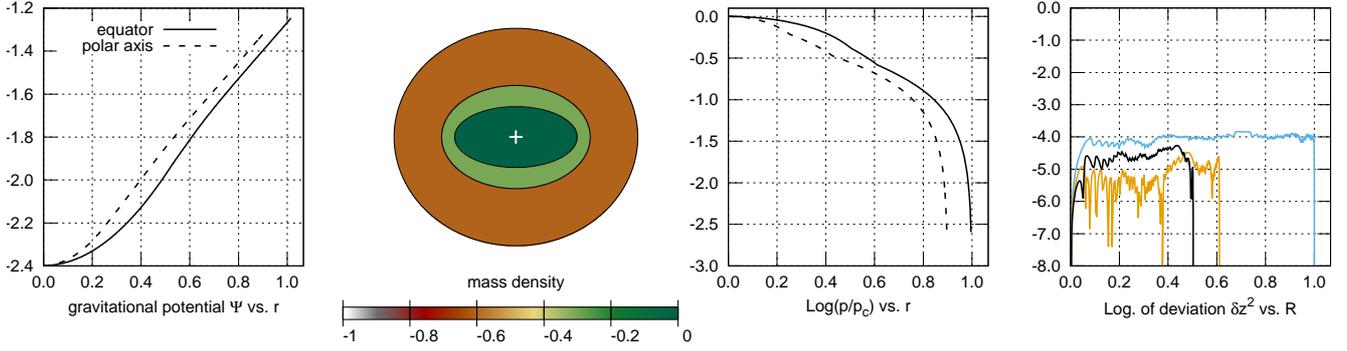}
\caption{Same legend as for Fig. \ref{fig:graphrhoA.ps} but for configuration B (confocal case). See Tab. \ref{tab:results-testB} for input/output parameters.}
\label{fig:graphrhoB.ps}
\end{figure*}
\begin{figure*}
 \includegraphics[height=17.5cm,bb=70 90 240 730,clip==, angle=-90]{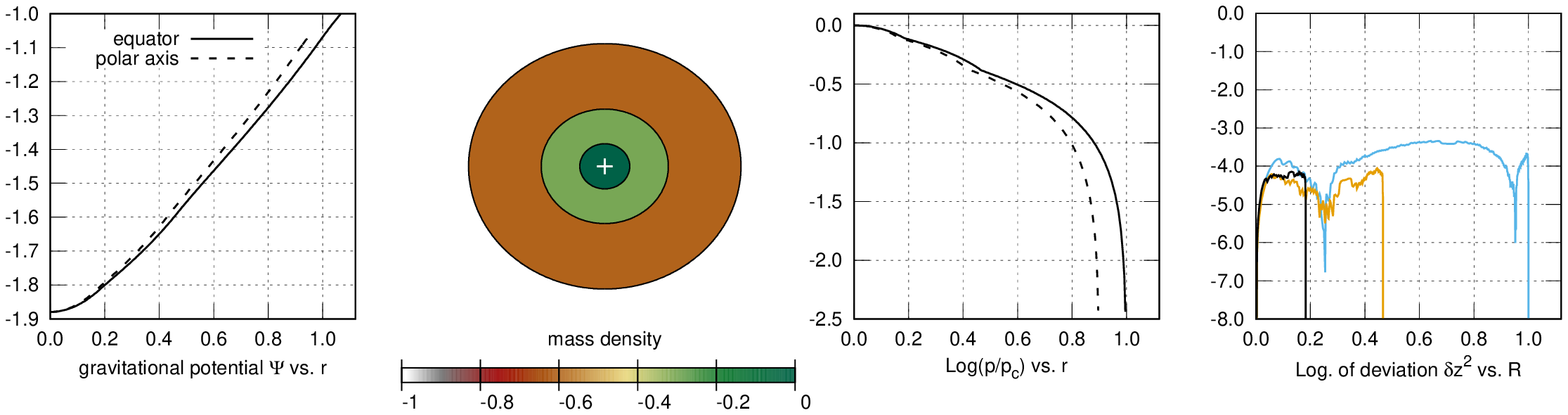}
\caption{Same legend as for Fig. \ref{fig:graphrhoA.ps} but for configuration C (coelliptical case). See Tab. \ref{tab:results-testC} for input/output parameters.}
\label{fig:graphrhoC.ps}
\end{figure*}
\begin{figure*}
 \includegraphics[height=17.5cm,bb=70 90 240 730,clip==, angle=-90]{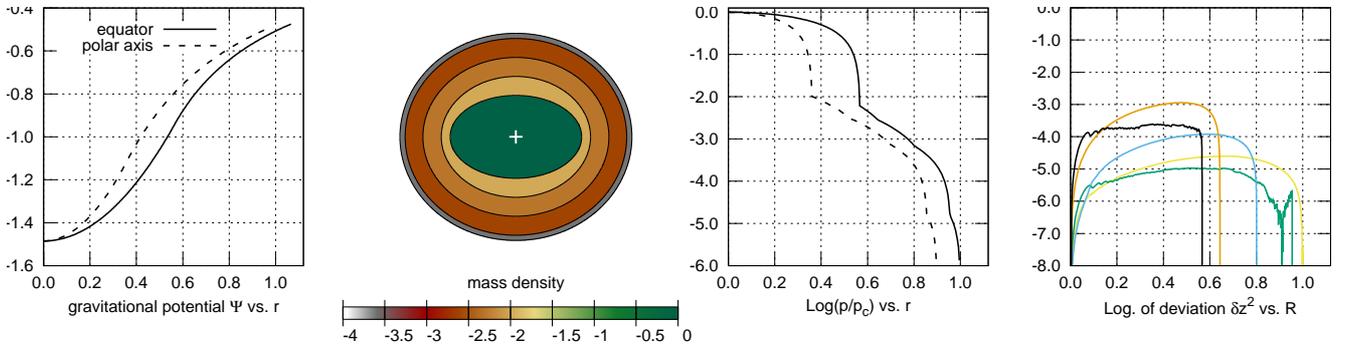}
\caption{Same legend as for Fig. \ref{fig:graphrhoA.ps} but for configuration D (dense, rapidly-rotating interior). See Tab. \ref{tab:resultsD} for input/output parameters.}
\label{fig:graphrhoD.ps}
\end{figure*}
\begin{figure*}
 \includegraphics[height=17.5cm,bb=70 90 240 730,clip==, angle=-90]{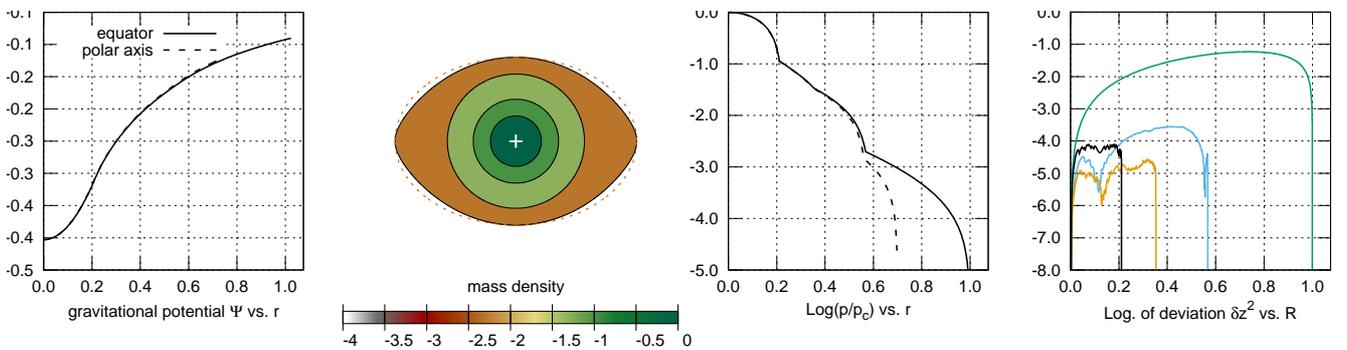}
\caption{Same legend as for Fig. \ref{fig:graphrhoA.ps} but for configuration E (dense, slowly-rotating interior). See Tab. \ref{tab:resultsE} for input/output parameters. The outer ellipse is shown ({\it red, dot line}).}
\label{fig:graphrhoE.ps}
\end{figure*}

\begin{table}
  \centering
  \begin{tabular}{lrrr}
    \multicolumn{4}{c}{{\bf configuration B} (input)}\\
   layer  & $i=1$ & $2$ & $3$  \\ \hline \hline
    $q_{i,3}\bar{\epsilon}_i$  & $0.25166$    &  $0.42725$  & $0.9$  \\
    $\alpha_i$                      & $2$ & $2$ &  \\ \hline \hline
 \end{tabular}\\\bigskip
  \begin{tabular}{lrrr}
    & {\tt DROP}-code & this work\\\hline
    $q_{1,3}$ & $0.50156$ & $0.50332$\\
    $q_{2,3}$ & $0.60936$ & $0.61036$\\
    $\epsilon_1$ & $0.86567$ & $0.86602$ \\
    $\epsilon_2$ & $0.71400$ & $0.71414$ \\
    $c_{1,3}$ & $-0.00061$ & $0$  \\     
    $c_{2,3}$ & $-0.00015$ & $0$  \\     
    $V/a_3^3$ & $3.77057$ & $3.76991$ \\
    $\pc/\pi G \rho_{3} a_{3}^2$  & $2.40553$ & $2.40491$\\ 
$p_1^*/\pi G \rho_{3} a_{3}^2$ &  $1.48071$ & $1.46845$\\
$p_2^*/\pi G \rho_{3} a_{3}^2$ &  $0.81826$ & $0.80963$\\
$\tilde{\Omega}_1^2$ & $0.76767$ & $0.76690$ & $0.71690^a$\\
$\tilde{\Omega}_2^2$ & $0.30654$ & $0.30623$ & $0.31381^a$ \\
$\tilde{\Omega}_3^2$ & $0.06867$ & $0.06859$ & $0.07510^a$ \\
    $M/\rho_3 a_{3}^3$  & $4.97044$ & $4.97076$\\
    $\nu_1$ & $0.21445$ & $0.21489$\\
    $\nu_2$ & $0.16102$ & $0.16081$\\ \hline
     \multicolumn{4}{l}{$^*$value on the polar axis}\\
    \multicolumn{4}{l}{$^a$from (\ref{eq:omegalapprox}) and (\ref{eq:omegajapprox})}\\
  \end{tabular} 
   \caption{Same legend as for Tab. \ref{tab:resultsA} but for configuration B (confocal case, ${\cal L}=3$) shown in Fig. \ref{fig:graphrhoB.ps}.}
  \label{tab:results-testB}
\end{table}

\begin{table}
  \centering
  \begin{tabular}{lrrr}
    \multicolumn{4}{c}{{\bf configuration C} (input)}\\
   layer  & $i=1$ & $2$ & $3$  \\ \hline\hline
    $q_{i,3}\bar{\epsilon}_i$  &  $0.165$  & $0.42$    &  $0.9$  \\
    $\alpha_i$                      & $2$ & $2$ & \\ \hline \hline
 \end{tabular} \\\bigskip
  \begin{tabular}{lrrr}
    & {\tt DROP}-code & this work\\\hline
    $q_{1,3}$ & $0.18309$ & $0.18333$\\
    $q_{2,3}$ & $0.46650$ & $0.46666$\\
    $\epsilon_1$ & $0.43346$ & $0.43588$\\
    $\epsilon_2$ & $0.43526$ & $0.43588$\\
    $c_{1,3}$  & $-0.18370$ & $-0.18361$ \\
    $c_{2,3}$  & $-0.14877$ & $-0.14862$\\
    $V/a_{\cal L}^3$ & $3.76893$ & $3.76991$ \\
   $\pc/\pi G \rho_{3} a_{3}^2$ & $1.46945$ & $1.46956$\\ 
$p_1^*/\pi G \rho_{3} a_{3}^2$  &  $1.15414$ & $1.15434$\\
$p_2^*/\pi G \rho_{3} a_{3}^2$  &  $0.62076$ & $0.62086$\\
$\tilde{\Omega}_1^2$ & $0.15008$ & $0.15004$ & $0.14559^a$\\
$\tilde{\Omega}_2^2$ & $0.09200$ & $0.09197$ & $0.08932^a$\\
$\tilde{\Omega}_3^2$ & $0.06533$ & $0.06531$ & $0.06338^a$\\
    $M/\rho_3 a_{3}^3$ & $4.19840$ & $4.19950$\\
    $\nu_1$ & $0.02213$ & $0.0223$\\
    $\nu_2$ & $0.17141$ & $0.17139$\\ \hline
     \multicolumn{4}{l}{$^*$value on the polar axis}\\
    \multicolumn{4}{l}{$^a$from (\ref{eq:omegalapprox}) and (\ref{eq:omegajapprox})}\\
  \end{tabular} 
   \caption{Same legend as for Tab. \ref{tab:resultsA} but for configuration C (coelliptical case, ${\cal L}=3$) shown in Fig. \ref{fig:graphrhoC.ps}.}
  \label{tab:results-testC}
\end{table}

\begin{table}
  \centering
  \begin{tabular}{lrrrrr}
    \multicolumn{6}{c}{{\bf configuration D} (input)}\\
   layer  & $i=1$ & $2$ & $3$ & $4$ & $5$  \\ \hline \hline
    $q_{i,5}\bar{\epsilon}_i$  &  $0.36$  & $0.525$  & $0.69$ & $0.855$ & $0.9$  \\
    $ \alpha_i$   & $10^2$ & $2$ & $2$ & $10$  \\ \hline \hline
 \end{tabular}\\\bigskip      
 \begin{tabular}{lrrr}
   & {\tt DROP}-code & this work\\\hline
   $\epsilon_1$ & $0.77201$ & $0.77211$ \\
   $\epsilon_2$ & $0.57804$ & $0.57804$ \\
   $\epsilon_3$ & $0.50905$ & $0.50906$  \\
   $\epsilon_4$ & $0.44510$ & $0.44510$ \\
    $q_{1,5}$ & $0.56639$ & $0.56650$\\
    $q_{2,5}$ & $0.64338$ & $0.64337$\\
    $q_{3,5}$ & $0.80163$ & $0.80164$\\
    $q_{4,5}$ & $0.95479$ & $0.95479$\\
    $c_{1,5}$  & $+0.00120$ & $+0.00132$\\
    $c_{2,5}$  & $-0.05168$ & $-0.05169$\\
    $c_{3,5}$  & $-0.023487$ & $-0.02346$\\
    $c_{4,5}$  & $-0.00938$ & $-0.00938$\\
    $V/a_{5}^3$ & $3.77000$ & $3.76991$ \\
    $\pc/\pi G \rho_{5} a_{5}^2 \times 10^{-6}$ & $1.92418$ & $1.92191$\\ 
$p_1^*/\pi G \rho_{5} a_{5}^2 \times 10^{-4}$ &  $1.95711$ & $1.95671$\\
$p_2^*/\pi G \rho_{5} a_{5}^2 \times 10^{-4}$ &  $0.58740$ & $0.58659$\\
$p_3^*/\pi G \rho_{5} a_{5}^2 \times 10^{-4}$ &  $0.15143$ & $0.15121$\\
$p_4^*/\pi G \rho_{5} a_{5}^2 \times 10^{-4}$ &  $0.00323$ & $0.00322$\\
$\tilde{\Omega}_1^2 \times 10^{-2}$ & $6.77290$ & $6.77090$ & $6.35370^a$\\
$\tilde{\Omega}_2^2 \times 10^{-2}$ & $0.67729$ & $0.67482$ & $0.82996^a$\\
$\tilde{\Omega}_3^2 \times 10^{-2}$ & $0.45152$ & $0.45059$ & $0.59140^a$\\
$\tilde{\Omega}_4^2 \times 10^{-2}$ & $0.30101$ & $0.30045$ & $0.40741^a$\\
$\tilde{\Omega}_5^2 \times 10^{-2}$ & $0.27365$ & $0.27317$ & $0.37262^a$\\
    $M/\rho_5 a_5^3 \times 10^{-3}$ & $1.98941$ & $1.98634$\\
    $\nu_1$             & $0.97462$ & $0.97453$\\
    $\nu_2 \times 10^2$ & $0.84950$ & $0.85855$\\
    $\nu_3 \times 10^2$ & $0.95480$ & $0.95361$\\
    $\nu_4 \times 10^2$ & $0.70774$ & $0.70862$\\ \hline
     \multicolumn{4}{l}{$^*$value on the polar axis}\\
    \multicolumn{4}{l}{$^a$from (\ref{eq:omegalapprox}) and (\ref{eq:omegajapprox})}\\
  \end{tabular} 
   \caption{Same legend as for Tab. \ref{tab:resultsA} but for configuration D (type-V solution, massive and rapidly rotating core) shown in Fig. \ref{fig:graphrhoD.ps}.}
   \label{tab:resultsD}
\end{table}

\begin{table}
  \centering
  \begin{tabular}{lrrrr}
    \multicolumn{5}{c}{{\bf configuration E} (input)}\\
   layer  & $i=1$ & $2$ & $3$ & $4$  \\ \hline\hline
    $q_{i,4}\bar{\epsilon}_i$  &  $0.21$  & $0.35$  & $0.56$ & $0.7$  \\
    $\alpha_i$    & $10$ & $2$ & $10$ \\ \hline \hline
  \end{tabular}\\\bigskip 
  \begin{tabular}{lrrr}
    & {\tt DROP}-code & this work\\\hline 
    $q_{1,4}$ & $0.21082$ & $0.21082$\\
    $q_{2,4}$ & $0.35325$ & $0.35327$\\
    $q_{3,4}$ & $0.56738$ & $0.56748$\\
    $\epsilon_1$  & $0.08837$ &  $0.08859$\\
    $\epsilon_2$  & $0.13546$ & $0.13594$\\
    $\epsilon_3$  & $0.16083$ &  $0.16184$\\
    $c_{1,4}$  & $-0.50965$ & $-0.50965$\\
    $c_{2,4}$  & $-0.50771$ & $-0.50769$\\
    $c_{3,4}$  & $-0.50167$ & $-0.50156$\\
    $V/a_{4}^3$ & $2.61819$ & $2.93215$ \\
    $\pc/\pi G \rho_{4} a_4^2 \times 10^{-3}$ & $1.33540$ & $1.33388$\\ 
$p_1^*/\pi G \rho_{4} a_{4}^2 \times 10^{-3}$ &  $0.15182$ & $0.15180$\\
$p_2^*/\pi G \rho_{4} a_{4}^2 \times 10^{-3}$ &  $0.04533$ & $0.04532$\\
$p_3^*/\pi G \rho_{4} a_{4}^2 \times 10^{-3}$ &  $0.00191$ & $0.00192$\\
$\tilde{\Omega}_1^2 \times 10^{-2}$ & $0.02265$ & $0.03742$ & $0.12355^a$\\
$\tilde{\Omega}_2^2 \times 10^{-2}$ & $0.22659$ & $0.21565$ & $0.25293^a$\\
$\tilde{\Omega}_3^2 \times 10^{-2}$ & $0.22659$ & $0.21308$ & $0.20310^a$\\
$\tilde{\Omega}_4^2 \times 10^{-2}$ & $2.26594$ & $2.26783$ & $1.41244^a$\\
    $M/\rho_4 a_{4}^3 $ & $18.28944$ & $18.59843$\\
    $\nu_1$ & $0.42828$ & $0.42045$\\
    $\nu_2$ & $0.15719$ & $0.15471$\\
    $\nu_3$ & $0.31262$ & $0.30778$\\\hline
     \multicolumn{4}{l}{$^*$value on the polar axis}\\
    \multicolumn{4}{l}{$^a$from (\ref{eq:omegalapprox}) and (\ref{eq:omegajapprox})}\\
  \end{tabular} 
   \caption{Same legend as for Tab. \ref{tab:resultsA} but for configuration E (type-V solution, slowly-rotating core) shown in Fig. \ref{fig:graphrhoE.ps}.}
   \label{tab:resultsE}
\end{table}

\subsection{A few tests. Efficiency and limit of the method}
\label{sec:tests}

As the first test, we consider the parameter set of Tab. \ref{tab:resultsA} (configuration A). The sequence of rotation rates is computed from (\ref{eq:omegalrecursionvector}) and (\ref{eq:omegajrecursionvector}) for $j=\{ 1,2,3\}$, and the pressure at the interfaces and at the centre is estimated by recursion from formula given in the Appendix \ref{sec:pressures}. As we make systematic comparisons with the numerical solutions obtained from the SCF-method \citep{bh21} which is used as ``the reference'', it is important that all the formulas are fed with the same parameters (see note \ref{note:drop}). For this configuration, the location of points A$_i$ on the polar axis, the mass-density jumps and the fractional radii $q_i$ are the values delivered by the {\tt DROP}-code on output. The results are reported in the same table  columns 2 and 3). We notice the remarkable agreement between the analytical approach and the numerical reference. The deviations observed for the pressures, rotation rates and fractional masses are of the order of a few $10^{-3}$ in relative.

The second and third tests illustrate the two cases considered in Sects. \ref{sec:confocal} and  \ref{sec:coell} respectively, with  ${\cal L}=3$ in both cases. We give in Tab. \ref{tab:results-testB} the data obtained for the confocal case. Here, the mass-density jumps and the ellipticities are the same for both methods. The fractional radii injected in the formula are easily computed from (\ref{eq:confocalc}). This is configuration B. The pressure, the mass density and the deviations between bounding surfaces are displayed in Fig. \ref{fig:graphrhoB.ps}. The parameter set and the results for the coelliptical confiuration are listed in Tab. \ref{tab:results-testC} and the associated plots are in Fig. \ref{fig:graphrhoC.ps}. The procedure is similar: the location of points A$_i$ and the mass-density jumps are imposed for both methods, which determines the unique ellipticity (this value is imposed by the outermost spheroid, namely $\bar{\epsilon}=$OA$_{\cal L}/$OB$_{\cal L}$). This is configuration C. Again, we see that the two techniques give very similar results with a precision better than $10^{-3}$ in relative on the main quantities. By changing slightly the input parameters, the code can deliver yet smaller confocal parameters.

In the fourth example, we consider a very dense (by a factor $10^2$), rapidly (by a factor $10$) rotating core surrounded with $4$ low-mass layers in relative motion. The parameters are given in Tab. \ref{tab:resultsD}. This is configuration D. The pressure, the mass density and the deviations between surfaces are given in Fig. \ref{fig:graphrhoD.ps}. The deepest layer is significantly flattened by rotation, as we have chosen $\bareone \approx 0.64$ (this value is below the threshold of dynamical stability for single body). The formulas are fed with the fractional radii and ellipticities output by the simulation. The mass-density jumps are the same. As the table shows (column 2 and 3), the numerical SCF-method and the actual formalism compare very well, at a precision level of a few $10^{-3}$ typically. The confocal parameters are all very small, and much smaller than in previous examples. Note that we are far from a situation with small ellipticities.

The last example is some kind of reverse situation. The axis ratio of the outermost layer is set to $0.7$, and its rotation rate is much larger than that of the dense core (by a factor $100$). As above, the characteristics of the surface computed from the code are used as parameters for the analytical approach. The input parameters and output data are gathered in Tab. \ref{tab:resultsE}. This is configuration E, which is displayed in Fig. \ref{fig:graphrhoE.ps}. The outer layer is especially flattenend by centrifugation while the core is very close to spherical. This example also shows the limit in the model. This is indicated by the confocal parameters $c_{i {\cal L}}$ which are quite ``large''. Besides, we see that the outer layer deviate significantly from a spheroid, namely by more than $10 \%$ beyond $R/a_{\cal L} \gtrsim 0.2$. The actual formalism, to be valid, not only requires $\cij \rightarrow 0$ but also that each bounding surface must be close to a spheroid, which property is not represented only by the $\qij$'s and $\epsilon_i$'s.

\begin{figure*}
 \includegraphics[height=17.5cm,bb=60 90 240 730,clip==, angle=-90]{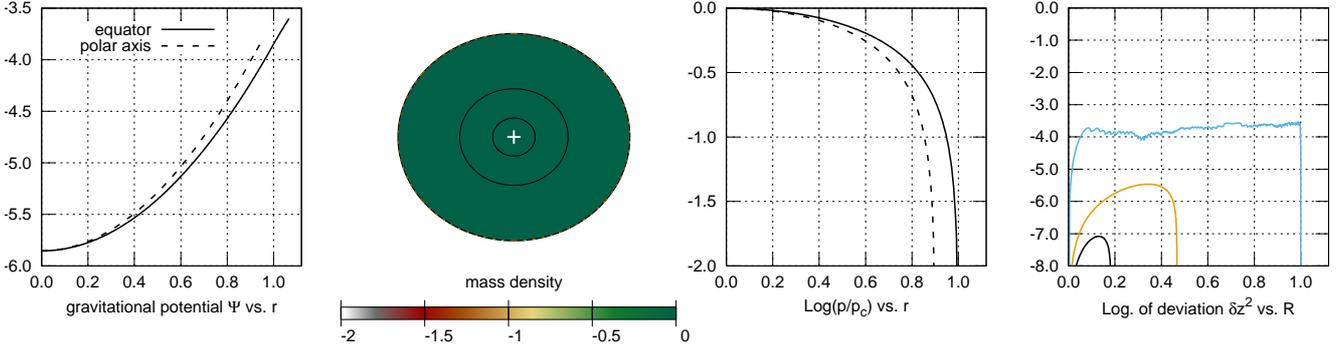}
\caption{Same legend as for Fig. \ref{fig:graphrhoA.ps} but for configuration G (type-C solution; coelliptical case). See Tab. \ref{tab:results-testG} for input/output parameters.}
\label{fig:graphrhoG.ps}
\end{figure*}

\section{Note on type-C solutions (global rotation)}
\label{sec:typec}

For ${\cal L}=2$ (see Paper I), when the pressure is constant all along the interface $E_1$ between the embedded spheroid and the host, then the two layers share the same rotation rate (solutions are then called ``type-C'' solutions). But the mass-density jump at $E_1$ is not arbitrary. In contrast, when the pressure varies with the cylindrical radius $R$ | the variation is quadratic for rigid rotations|, then $\Omega$ differs from one layer to other. We have in this case ``type-V'' solutions. This is possible only if a mass-density jump $\alpha_1 \ne 1$ is present at the interface; see also \cite{bh21}. These two classes of solutions also exist for ${\cal L}>2$. Actually, the series of rotation rates obtained from  (\ref{eq:omega1recursionvector}) do not assign a unique value to all $\tilde{\Omega}_i'$ s, unless a ``fortunate'' selection of the input parameters. Each layer has therefore its own rotation rate $\Omega_i$. In this sense, (\ref{eq:omega1recursionvector}) delivers type-V solutions in general. When rotation is global, i.e. $\tilde{\Omega}_i= \tilde{\Omega}$ for all $i$, then we have from (\ref{eq:omegalrecursionvector}) and (\ref{eq:omegajrecursionvector}), assuming $\alpha_i -1 \ne 0$
\begin{flalign}
  \label{eq:omegatypec}
  &\tilde{\Omega}^2=-  {\cal M} (\epsj) \, \mathbf{X} \cdot \left[ \mathbf{P}({\rm B}_j) + \mathbf{C}({\rm A}_j,{\rm B}_j)\right], \quad j \in [1,{\cal L}].
  \end{flalign}
It follows that the mass density jumps $\alpha_i$, the ellipticities $\epsilon_i$ and the fractional sizes $\qij$ are linked together. We see that $\tilde{\Omega}^2$ can be eliminated, for instance by subracting each equation by the last one. We then get a {\it linear system of ${\cal L}-1$ equations}, where the first ${\cal L}-1$ components $X_i$ of vector $\mathbf{X}$ are the unknown ($\epsilon_i$ and $\qij$ are fixed in the procedure). This system of equations takes the form
\begin{flalign}
  \label{eq:dxe}
\sum_{i=1}^{{\cal L}-1}{D_{i,j} X_j}-D_{{\cal L}, j}=0, \quad j\in [{\cal L}-1],
\end{flalign}
where the matrix elements are given by
\begin{flalign}
  \label{eq:dij}
  &D_{i,j}= {\cal M} (\epsj)\left[ P_i({\rm B}_j) + C_i({\rm A}_j,{\rm B}_j) \right]\\
  \nonumber
  & \qquad \qquad-{\cal M} (\epsilon_{\cal L})\left[ P_i({\rm B}_{\cal L}) + C_i({\rm A}_{\cal L},{\rm B}_{\cal L}) \right],
\end{flalign}
and the second member is obtained for $i={\cal L}$, namely
\begin{flalign}
  \label{eq:ei}
  &D_{{\cal L}, j}= {\cal M} (\epsj)\left[ P_{\cal L}({\rm B}_j) + C_{\cal L}({\rm A}_j,{\rm B}_j) \right]\\
  \nonumber
  & \qquad \qquad -{\cal M} (\epsilon_{\cal L})\left[ P_{\cal L}({\rm B}_{\cal L}) + C_{\cal L}({\rm A}_{\cal L},{\rm B}_{\cal L}) \right].
\end{flalign}
This system is easily solved numerically by standard techniques. The mass-density jumps associated with the solution $\mathbf{X}$ are denoted $\alpha_{Ci}$ in the following.

\begin{table}
  \centering
  \begin{tabular}{lrrrr}
    \multicolumn{5}{c}{{\bf configuration F} (input)}\\
   layer & $i=1$ & $2$ & $3$ & $4$ \\ \hline \hline
    $q_{i,4}\bar{\epsilon}_i$  & $0.21$    &  $0.45$   & $0.72$   &  $0.9$  \\ \hline \hline
   \end{tabular}\\\bigskip
  \begin{tabular}{lrr}
    & {\tt DROP}-code & this work\\\hline
    $q_{1,4}$ & \multicolumn{2}{c}{$0.22176$} \\
    $q_{2,4}$ & \multicolumn{2}{c}{$0.48230$} \\
    $q_{3,4}$ & \multicolumn{2}{c}{$0.78690$} \\
    $\epsilon_1$ & \multicolumn{2}{c}{$0.32141$} \\
    $\epsilon_2$ & \multicolumn{2}{c}{$0.35984$} \\
    $\epsilon_3$ & \multicolumn{2}{c}{$0.40349$} \\ 
    $c_{1,4}$ & \multicolumn{2}{c}{$-0.18491$} \\
    $c_{2,4}$ & \multicolumn{2}{c}{$-0.15987$} \\
    $c_{3,4}$& \multicolumn{2}{c}{$-0.08918$}  \\ 
    $\alpha_1$ & $2$ & $1.92825$ \\
    $\alpha_2$ & $2$ & $2.02834$ \\
    $\alpha_3$ & $2$ & $1.97393$ \\
    $V/a_{\cal L}^3$ & $3.76482$ & $3.76991$ \\
    $\pc/\pi G \rho_{4} a_{4}^2$    & $6.22120$ & $6.01587$\\ 
    $p_1^*/\pi G \rho_{4} a_{4}^2$ &  $4.23392$ & $4.16533$\\
    $p_2^*/\pi G \rho_{4} a_{4}^2$  &  $1.89159$ & $1.85745$\\
    $p_3^*/\pi G \rho_{4} a_{4}^2$  &  $0.45253$ & $0.44881$\\
    $\tilde{\Omega}_1^2$ & $0.13002$ & $0.12900$\\
    $\tilde{\Omega}_2^2$ & $0.13002$ & $0.12900$ \\
    $\tilde{\Omega}_3^2$ & $0.13002$ & $0.12900$\\
    $\tilde{\Omega}_4^2$ & $0.13002$ & $0.12900$\\
    $M/\rho_4 a_{4}^3$ & $6.68126$ & $6.68742$ \\
    $\nu_1$ & $0.05178$ & $0.05175$ \\
    $\nu_2$ & $0.23664$ & $0.23639$ \\
    $\nu_3$ & $0.42743$ & $0.42737$ \\ \hline
    \multicolumn{3}{l}{$^*$value on the polar axis}
  \end{tabular} 
   \caption{Same legend as for Tab. \ref{tab:resultsA} but for configuration F (type-C solution) associated with the same geometrical parameters as for configuration A; see Tab. \ref{tab:resultsA}. The mass-density jumps are found by solving (\ref{eq:dxe}), which ensures that all layers rotates at the same rate.}
   \label{tab:resultsF}
\end{table}

\begin{table}
  \centering
  \begin{tabular}{lrrr}
    \multicolumn{4}{c}{\bf configuration G}\\
   layer  & $i=1$ & $2$ & $3$  \\ \hline\hline
    $q_{i,3}\bar{\epsilon}_i$  &  $0.165$  & $0.42$    &  $0.9$  \\\hline \hline
  \end{tabular}\\\bigskip
  \begin{tabular}{lrrr}
    & {\tt DROP}-code & this work\\\hline
    $q_{1,3}$ & $0.18334$ & $0.18333$\\
    $q_{2,3}$ & $0.46669$ & $0.46666$\\
    $\epsilon_1$ & $0.43599$ & $0.43588$\\
    $\epsilon_2$ & $0.43600$ & $0.43588$\\
    $c_{1,3}$  & $-0.18361$ & $-0.18361$ \\
    $c_{2,3}$  & $-0.14859$ & $-0.14862$\\
    $V/a_{\cal L}^3$ & $3.77310$ & $3.76991$ \\
    $\pc/\pi G \rho_{3} a_{3}^2$ & $0.58626$ & $0.58615$\\ 
    $p_1^*/\pi G \rho_{3} a_{3}^2$  &  $0.56656$ & $0.56645$\\
    $p_2^*/\pi G \rho_{3} a_{3}^2$  &  $0.45861$ & $0.45850$\\
    $\tilde{\Omega}$ & $0.05208$ & $0.05202$ \\
    $M/\rhol a_{\cal L}^3$ & $3.77311$ & $3.76991$\\
    $\nu_1$ & $0.00641$ & $0.00616$\\
    $\nu_2$ & $0.09560$ & $0.09546$\\ \hline
    \multicolumn{3}{l}{$^*$value on the polar axis}
  \end{tabular} 
   \caption{Same legend as for Tab. \ref{tab:resultsA} but for configuration G (type-C solution) shown in Fig. \ref{fig:graphrhoG.ps}, associated with the same geometrical parameters as for configuration C; see Tab. \ref{tab:results-testC}.}
  \label{tab:results-testG}
\end{table}

As examples of type-C solutions, we have reconsidered configurations A, B and C discussed in Sect. \ref{sec:tests} by conserving the same geometrical parameters, namely the fractional radii $q_{i {\cal L}}$ and ellipticipties $\epsilon_i$, while the mass-density jumps are deduced from (\ref{eq:dxe}). For parameters of configuration A (see Tab. \ref{tab:resultsA}), we find $X_1 \approx 3.71658$, $X_2 \approx 2.03476$ and $X_3 \approx 0.97393$, which yields $\alpha_{C1} \approx 1.92825$, $\alpha_{C2} \approx 2.02834$ and $\alpha_{C3} \approx 1.97393$. We call this new configuration F. The output quantities are reported in Tab. \ref{tab:resultsF}. Regarding parameters of configuration B (confocal state), (\ref{eq:dxe}) admits $X_1 \approx 2.65560$ and $X_2 \approx -2.77502$ as the solution, which leads to $\alpha_{C1} \approx 0.88057$ and $\alpha_{C2} \approx -1.77502$. This is therefore not a relevant physical state. For parameters of configuration C (coelliptical case), we find ${\mathbf X}={\mathbf 0}$, which means that all layers have the same mass density, and are therefore indistinguishable (this is a single body). We call this new configuration G. The results are listed in Tab. \ref{tab:results-testG}. The structure and the deviations between surfaces are displayed in Fig. \ref{fig:graphrhoG.ps}

\section{Approximations for small ellipticities}
\label{sec:lowrotation}

Ellipticities close to zero are generally associated with states of low rotation. As in Paper I, we can expand each component of $\mathbf{P}({\rm B}_j)$ in $\epsilon_i \ll 1$. It is easy to show that
\begin{flalign}
 {\cal M}(x){\cal P}(x,y) \approx -\frac{2}{3}\left(x^2-\frac{3}{5}y^2\right),
\end{flalign}
for $x \ll 1$ and $y \ll 1$. So, if we define $\sqrt{k_{ij}}=\epsi/\epsj$, the components of $ {\cal M}(\epsj){\mathbf P}({\rm B}_j)$ can be approximated by
\begin{flalign}
  {\cal M}(\epsj){\rm P}_i({\rm B}_j) \approx \frac{2}{15} \times
\begin{cases}
\epsi^2 (3 \qij^2 -5 \kji)\qij^3,\\ \qquad \qquad \text{for } i < j,\\\\
\epsj^2(- 5 +3 \kij), \quad \text{for } i \ge j.
\end{cases}
\end{flalign}
This separation comes directly from (\ref{eq:pvect}). Actually, we have $\mathbf{P}_i({\rm B}_j)=f_{i,j}{\cal P}(\epsj,\epsilon_{i,j}')$ for $i<j$, while this component becomes ${\cal P}(\epsj,\epsi)$ for $i \ge j$. Under these conditions, and omitting the first-order correction represented by vector ${\mathbf C}$, (\ref{eq:omegalrecursionvector}) becomes
\begin{flalign}
  \nonumber
  &\tilde{\Omega}_{\cal L}^2 \approx  \frac{4}{15}\epsilon_{\cal L}^2\\
  & \qquad -\frac{2}{15}\sum_{i=1}^{{\cal L}-1}\epsi^2 \trho_{i+1}(\alpha_i-1)\left(3 q_{i, {\cal L}}^2 -5 k_{{\cal L},i}\right) q_{i {\cal L}}^3.
    \label{eq:omegalapprox}
\end{flalign}
For internal layers $j \in [1,{\cal L}-1]$, we use (\ref{eq:omegajrecursionvector}). If we separate the summation into two parts (see Sect. \ref{subsec:pot}), we have 
\begin{flalign}
  \label{eq:omegajapprox}
  &\alpha_j\tilde{\Omega}_j^2 \approx \tilde{\Omega}_{j+1}^2 \\
  \nonumber
  & \qquad - (\alpha_j-1)\frac{2}{15} \left[ \sum_{i=1}^{j-1}\epsi^2 \trho_{i+1}(\alpha_i-1)\left(3 \qij^2 -5 k_{j,i}\right) \qij^3\right.\\
  \nonumber
 & \qquad \qquad  \left.  + \sum_{i=j}^{{\cal L}-1}\epsj^2 \trho_{i+1}(\alpha_i-1)\left( 3 \kij -5 \right) + \epsj^2 \left( 3 k_{{\cal L},j} -5 \right)  \right],
  \end{flalign}
where the last term inside the brakets corresponds to $i={\cal L}$, i.e. the last components in (\ref{eq:xvect}) and (\ref{eq:pvect}). Note that, for $j=1$, the first summation is absent. We give in Tabs. \ref{tab:resultsA} to \ref{tab:resultsE} the series of rotation rates computed from (\ref{eq:omegalapprox}) and (\ref{eq:omegajapprox}) for configurations A to E considered above. We see that these zero-order formula are sufficient to get a precision of the order of a few percents. Unsurprisingly, it fails when the confocal parameters are not small in absolute compared to unity (this assumption underlies the present formalism as a whole), which is clearly the case of configuration E. Configurations, like examples D and E, with large mass-density jumps are also poorly reproduced. This is inherent in the formula where the $\alpha_i$'s magnify the errors.

\section{Summary and perspectives}

In this article, we have investigated the conditions under which a self-gravitating structure made of ${\cal L}$ nested layers in relative rotation be in equilibrium (under axial and equatorial symmetries). It generalizes the approach considered in Paper I focused on the two-layer case. There are three main assumptions, namely: i) all layers are separated by concentric, spheroidal (non-intersecting) surfaces, ii) each layer is homogeneous, and iii) each layer can rotate at its own rate. Assumptions i) and ii) enable to use exact results from potential theory \citep{chandra69,binneytremaine87}. We have shown that the problem admits approximate solutions compatible with rigid rotation provided confocal parameters $\cij$ are close to zero (in the limit $\cij \rightarrow 0$, these solutions are exact, according to Poincar\'e's theorem; see the Introduction). Then, the full sequence of rotation rates $\Omega_1, \dots \Omega_{\cal L}$ can be generated from recursion; see (\ref{eq:omegalrecursionvector}) and (\ref{eq:omegajrecursionvector}). A similar procedure leads to the interface pressures along the polar axis and to the central pressure; see the Appendix \ref{sec:pressures}.  There are $3 {\cal L}-2$ input parameters in total, namely the ellipticities $\epsilon_i \in [0,1]$, the fractional radii $q_{i, {\cal L}} \in [0,1]$ and the mass-density jumps $\alpha_i \in [1,\infty[$ at the interfaces (other options are possible). The approach is vectorial, totally scale free, and is easily implemented in practice; see, for instance, the Appendix \ref{sec:F90} for a basic Fortran 90 code. As discussed, any set of input parameters does not necessarily lead to a physical solution. Actually, all the $\Omega^2_i$'s must be positive values, which is difficult to guarantee without considering numbers in the formulas (the number of input parameters determines the size of the parameter space). Nevertheless, we have qualitatively shown that the most favorable conditions for getting solutions are met for a positive ellipticity gradient outward (the layer are more and more oblate from centre to surface, at the opposite of confocal configurations). In general, the physical solution corresponds to layers in relative motion (i.e. type-V solutions), unless some specific choices of the input parameters. States where all rotation rates coincide (rotation is global; type-C solutions) are directly accessible by solving a linear system of ${\cal L}-1$ equations, where the mass-density jumps become the unknowns (the $\epsilon_i$'s and the $q_{i, {\cal L}}$'s are still parameters). Global rotation is therefore more constrained (and less numerous) than asynchronous states. In particular, we have demonstrated that confocal configurations are not possible in the presence of a negative mass-density gradient outward (no density inversion). This reinforces the conclusion by \cite{hamy90} and \cite{mmc83}. In a similar manner, we have shown that coelliptical configurations are not possible for global rotation, which prolongates Hamy's theorem limited to small ellipticities. Of particular interest is the case of small ellipticities, which applies to slowly-rotating stars and planets. In this purpose, the recursion formula, expanded in $\epsi^2 \ll 1$, takes a simple form. The formalism has been widely tested through examples by using the numerical SCF-method as the reference. It turns out that, provided the confocal parameters are small enough, namely $\cij^2 \ll 1$, then the precision of the method easily reaches $10^{-3}$ in relative, which should be sufficient for most applications.

    The formalism can help in better undertanding the internal structure (number of representative layers, ellipticities, fractional radii and mass-densities) of stars and planets from observational data (mass, shape, rotation rate, gravitational moments, etc.). Note that many models of planetary interiors are based on coelliptical configurations and global rotation \cite[e.g.][]{schu11,hub13,nettelmann2017,dc18}, which are clearly forbidden; see also \cite{cmm17}. By releasing the constraints in the ellipticities of layers \citep[e.g.][]{zhang96} | more drastically, in their shape \citep{net21} |, and by allowing different rotation rates \citep[e.g.][]{UrielCisnerosParra2020}, the observational data might be easily reproduced by a very limited number of homogeneous layers. An application to Jupiter is currently underway.
    
The paper opens onto a few exciting (eventually old) problems. This is, for instance, the determination of equilibrium sequences for various equations of states (like polytropic ones, for instance). Actually, in the limit ${\cal L} \rightarrow \infty$, the relationship between the $\alpha_i$'s, the series of ellipticities, fractional mass, and the polytropic index should be acccessible. Then, a connection with the solution of the Lane-Emden equation in the limit $\epsilon_i \rightarrow 0$ and solution might be also established. Another question concerns the assumption of non-intersecting surfaces, as considered for instance by \cite{ak74} and \cite{cai86}. Which conditions are required to get polar caps or equatorial caps ? This may be interesting in the context of exoplanets and ocean planets. It could also be very interesting to consider the existence of nested figures involving prolate spheroidal surface, which may occur, for instance, in the presence of circulation of magnetic field \citep{ktye11,fe14}. As shown here, rapidly rotating layers tend to produce quasi-spherical cores (see, for instance, configuration E).

\section*{Data availability} All data are incorporated into the article.

\section*{Acknowledgements}

I am grateful to the referee for his very precise report, and in particular for checking all the formulas, which has permitted to remove several typos. I would like to thank E. Di Folco for many interesting discussions about applications, B. Basillais and C. Staelen for testing the recurrence formulas in practical cases.

\bibliographystyle{mn2e}


\appendix

\noindent

\section{Interface pressures on the polar axis}
\label{sec:pressures}

As for the rotation rates, we proceed from the top layer down to the deepest one. The pressure on the polar axis is easily found since there is no centrifugal force at $R=0$. The pressure at the poles are {\bf denoted} $p_i({\rm A}_i) \equiv p_i^*$. By using (\ref{eq:bernoulli}) for a given layer $j+1$ at $E_j$ (bottom surface) and $E_{j+1}$ (top surface), we get
\begin{flalign}
  \frac{1}{\rho_{j+1}}\left(p_{j+1}^* -p_j^*\right) +  \Psi(E_{j+1},0)- \Psi(E_j,0)=0,
\end{flalign}
for $i=1,{\cal L}-1$. We have $p_{\cal L}=0$ along $E_{\cal L}$ and in particular at point A$_L$, which initializes the recursivity. By using (\ref{eq:psitotbis}), the above formula therefore reads
\begin{flalign}
  \begin{cases}
    p_{\cal L}^*=0,\\
   p_j^*  = p_{j+1}^* + \rho_{j+1}\left[\Psi(E_{j+1},0)- \Psi(E_j,0)\right],
  \end{cases}
\end{flalign}
with $j\in[1,{\cal L}-1]$, and
\begin{flalign}
  & \frac{\Psi(E_j,0)}{-\pi G a_j^2}= \sum_{i=1}^{{\cal L}-1} \left(\rho_i-\rho_{i+1}\right) \left\{ \fij \left[ A_0(\epsijprim)\xij \right. \right.\\
  & \left. \left.  \left. -A_3(\epsijprim)\barej^2 \right] \right\}\right|_{{\rm A}_j} +\rho_{\cal L} \left[ A_0(\epsilon_{{\cal L}, j}')x_{{\cal L}, j} -  A_3(\epsilon_{{\cal L}, j}') \barej^2\right].
  \nonumber
\end{flalign}
The central pressure $\pc$ is obtained in a similar way, from
\begin{flalign}
   \pc  = p_1^* + \rho_1\left[\Psi(E_1,0)- \psic\right],
\end{flalign}
where $\psic$ is the potential at the origin of coordinates which is found from (\ref{eq:psitot}), namely
\begin{flalign}
  \label{eq:psix}
  & \frac{\psic}{-\pi G a_{\cal L}^2}= \sum_{i=1}^{{\cal L}-1} \left(\rho_i-\rho_{i+1}\right) \fij A_0(\epsijprim)\xij \\
  \nonumber
  & \qquad\qquad\qquad +\rho_{\cal L} A_0(\epsilon_{{\cal L}, j}')x_{{\cal L} j}.
\end{flalign}
Note that the mass density jumps can easily be introduced by dividing this expression by $\rho_{\cal L}$
\section{The two-layer case}
\label{sec:twolayer}

For ${\cal L}=2$, (\ref{eq:xvect}), (\ref{eq:pvect}) and (\ref{eq:cvect}) reduce to
\begin{flalign}
  \mathbf{X} =
	\begin{pmatrix} 
		\alpha_1-1 \\ 
                1
	\end{pmatrix},
\end{flalign}
\begin{flalign}
   \mathbf{P}({\rm B}_j) = 
	\begin{pmatrix} 
		\left. f_{1j}{\cal P}(\epsj,\epsilon_{1, j}')\right|_{{\rm B}_j} \\ 
                {\cal P}(\epsj,\epsilon_2)
	\end{pmatrix},
\end{flalign}
and
\begin{flalign}
   \mathbf{C}({\rm A}_j,{\rm B}_j) = 
	\begin{pmatrix} 
		\left. f_{1,j} {\cal C}(\epsj,\epsilon_{1,j}')\right|_{{\rm A}_j}^{{\rm B}_j} \\ 
                0
	\end{pmatrix},
\end{flalign}
where $j \in \{1,2\}$. Then, for $j=2$ (i.e. the outer layer), we have ${\cal P}(\epsilon_2,\epsilon_2)=-1$, and (\ref{eq:omegalrecursionvector}) leads to
\begin{flalign}
  \nonumber
  &\tilde{\Omega}_2^2 = {\cal M}(\epsilon_2) \left\{1-(\alpha_1-1) \left[ \left. f_{1,2}{\cal P}(\epsilon_2,\epsilon_{1,2}') \right|_{B_2} \right. \right. \\
  &\left.\left. \qquad \qquad \qquad \qquad + \left.  f_{1,2} {\cal C}(\epsilon_2,\epsilon_{1,2}')\right|_{{\rm A}_2}^{{\rm B}_2} \right] \right\},
\end{flalign}
where $\epsilon_{1,2}'$ is to be replaced by $q_{1,2}\epsilon_{1}$ at point B$_2$, and by $\frac{q_{1,2}\epsilon_{1}}{\sqrt{1-c_{1,2}}}$ at point A$_2$. For the embedded ellipsoid, we set $j=1$ in (\ref{eq:omegajrecursionvector}), which leads to
\begin{flalign}
  &\alpha_1\tilde{\Omega}_1^2 = \tilde{\Omega}_2^2+ (\alpha_1-1) {\cal M} (\epsilon_1) \left[ \alpha_1-1 -{\cal P}(\epsilon_1,\epsilon_2) \right].
  \end{flalign} 
Note the absence of term ${\cal C}$ in this last relationship since $E_1$ is interior to $E_2$ and it is confocal with itself. We therefore recover the expressions reported in Paper I. It can be shown that (\ref{eq:omegalapprox}) and (\ref{eq:omegajapprox}) are also on accordance.

\onecolumn

\section{F90 program}
\label{sec:F90}

Not optimized. Routines {\tt cteA0(e)}, {\tt cteA1(e)}, and {\tt cteA3(e)} are required (see Paper I).

\begin{verbatim}
Program nsfoel
  ! gfortran nsfoel.f90; ./a.out
  Implicit None
  Integer,Parameter::AP=Kind(1.00E+00),NLAYER=4
  Real(Kind=AP),Parameter::PI=Atan(1._AP)*4
  Real(Kind=AP),dimension(1:NLAYER)::alpha,e,ebar,q,om2
  ! Statements
  ! input parameters L, qi's, ei's, alphai's (setup for configuration A)
  q(1:NLAYER)=(/2.2176752446908443E-01_AP,4.8230874759281428E-01_AP,7.8690212732359022E-01_AP,1._AP/)
  alpha(1:NLAYER)=(/2._AP,2._AP,2._AP,1._AP/)
  ebar(1:NLAYER)=(/0.21_AP,0.45_AP,0.72_AP,0.9_AP/)/q(1:NLAYER)
  e(1:NLAYER)=Sqrt(1._AP-ebar(1:NLAYER)**2)
  print*,"Ellipticities EPSILON(1:L)",e(1:NLAYER)
  print*,"Fractional sizes q(1:L-1)",q(1:NLAYER-1)
  print*,"Mass density jumps ALPHA(1:L-1)",alpha(1:NLAYER-1)
  ! TYPE-V SOLUTION
  ! input: NLAYER,e(1:NLAYER),q(1:NLAYER),alpha(1:NLAYER)
  ! output: om2(1:NLAYER)
  Call NFEL_typeV_Solution(NLAYER,e(1:NLAYER),q(1:NLAYER),alpha(1:NLAYER),om2(1:NLAYER))

Contains
  
  Subroutine NFEL_typeV_Solution(L,e,q,alpha,W2)
    Implicit none
    Integer,Intent(In)::L
    Real(Kind=AP),Dimension(1:L),Intent(In)::e,q,alpha
    Real(Kind=AP),Dimension(1:L),Intent(Out)::W2
    ! local
    Integer::I,J
    Real(Kind=AP),Dimension(1:L)::e2,ebar,rho,fvol,Avec,pif,psipole
    Real(Kind=AP),Dimension(1:L,1:L)::cij,xija,xijb,fija,fijb,eprimija,eprimijb,MPvec,MCvec
    Real(Kind=AP)::qij,mass,pc,psic
    ! initializations Ai's, rhoi's, qij's, xij's, fij's, e'ij's
    print*,"TYPE-V SOLUTION"
    e2(1:L)=e(1:L)**2;ebar(1:L)=Sqrt(1._AP-e2(1:L))
    rho(L)=1._AP;Avec(L)=1._AP
    Do I=L,1,-1
       If (I<L) then
          rho(I)=alpha(I)*rho(I+1);Avec(I)=rho(I)-rho(I+1)
       Endif
       Do J=1,L
          qij=q(I)/q(J)
          If (J>I) Then
             cij(I,J)=qij**2*e2(I)-e2(J)
             xijb(I,J)=1._AP;eprimijb(I,J)=qij*e(I)/Sqrt(xijb(I,J))
             fijb(I,J)=qij**3*ebar(I)/xijb(I,J)/sqrt(xijb(I,J)-qij**2*e2(I))
             xija(I,J)=1._AP+cij(I,J);eprimija(I,J)=qij*e(I)/Sqrt(xija(I,J))
             fija(I,J)=qij**3*ebar(I)/xija(I,J)/sqrt(xija(I,J)-qij**2*e2(I))
          Else
             cij(I,J)=0._AP
             xija(I,J)=qij**2;fija(I,J)=1._AP;eprimija(I,J)=e(I)
             xijb(I,J)=qij**2;fijb(I,J)=1._AP;eprimijb(I,J)=e(I)
          Endif
       Enddo
    Enddo
    print*,"Mass densities RHO(1:L)",rho(1:L)
    print*,"Confocal param. c(1:L-1,L)",cij(1:L-1,L)
    ! continued, Ai and M(ej).[Pi(Bj)+Ci(Aj,Bj)]
    psic=0._AP
    Do J=1,L
       psipole(J)=-fija(L,J)*(cteA0(eprimija(L,J))*xija(L,J)-cteA3(eprimija(L,J))*(1._AP-e2(J)))
       Do I=1,L-1
          psipole(J)=psipole(J)&
                  &-Avec(I)*fija(I,J)*(cteA0(eprimija(I,J))*xija(I,J)-cteA3(eprimija(I,J))*(1._AP-e2(J)))
          If (J>I) Then
             MPvec(I,J)=fijb(I,J)*(cteA3(eprimijb(I,J))*(1._AP-e2(J))-cteA1(eprimijb(I,J)))
             MCvec(I,J)=fijb(I,J)*(cteA0(eprimijb(I,J))*xijb(I,J)-(1._AP-e2(J))*cteA3(eprimijb(I,J)))&
                  &-fija(I,J)*(cteA0(eprimija(I,J))*xija(I,J)-(1._AP-e2(J))*cteA3(eprimija(I,J)))
          Else
             MPvec(I,J)=cteA3(e(I))*(1._AP-e2(J))-cteA1(e(I));MCvec(I,J)=0._AP
          Endif
       Enddo
       psic=psic-Avec(J)*cteA0(e(J))*q(J)**2
       psipole(J)=psipole(J)*q(J)**2
    Enddo
    MPvec(L,1:L)=cteA3(e(L))*(1._AP-e2(1:L))-cteA1(e(L));MCvec(L,1:L)=0._AP
    print*,"Interface potential PSI*(L:1) and central value PSIc",psipole(L:1:-1),psic
    ! sequence of rotation rates, TYPE-V SOLUTION
    J=L
    pif(J)=0._AP
    W2(J)=-dot_product(Avec(1:L),MPvec(1:L,J)+MCvec(1:L,J))
    Do J=L-1,1,-1
       fvol(J+1)=ebar(J+1)*q(J+1)**3-q(J)**3*ebar(J)
       W2(J)=(W2(J+1)-dot_product(Avec(1:L),MPvec(1:L,J)+MCvec(1:L,J))*(alpha(J)-1._AP))/alpha(J)
       pif(J)=rho(J+1)*(psipole(J+1)-psipole(J))+pif(J+1)
    Enddo
    pc=rho(1)*(psipole(1)-psic)+pif(1)
    print*,"Interface pressures p*(L:1) and central value pc",pif(L:1:-1),pc
    fvol(1)=q(1)**3*ebar(1)
    print*,"Rotation rates W2(1:L) (norm.)",W2(1:L)
    mass=dot_product(fvol(1:L),rho(1:L))*PI*4/3
    print*,"Total mass M",mass
    print*,"Fractional masses NU(1:L)",fvol(1:L)*rho(1:L)/mass*PI*4/3
  End Subroutine NFEL_typeV_Solution

End Program nsfoel
\end{verbatim}

\end{document}